\documentclass[a4paper]{article}

\usepackage[left=1.3in, right=1.3in, top=1in, bottom=1in]{geometry}

\usepackage{graphicx} 
\usepackage{float}
\usepackage{longtable}
\usepackage{caption}
\usepackage{subcaption}
\usepackage{ragged2e}
\usepackage[table]{xcolor}
\usepackage{changepage}
\usepackage{makecell}
\usepackage{amsfonts}
\usepackage[utf8]{inputenc}
\usepackage{newunicodechar}
\usepackage{siunitx} 
\usepackage{hyperref}
\usepackage{tcolorbox}
\usepackage[style=numeric, sorting=none, natbib=true, backend=biber, maxbibnames=100, maxcitenames=100, giveninits=true]{biblatex}
\usepackage{multirow}
\usepackage[title]{appendix}
\usepackage{titlesec}
\usepackage{booktabs}
\usepackage{array}
\title{\textbf{Machine Learning for RNA-Targeting Drug Design}}

\author{
  \textbf{Wissam Karroucha}$^{1,2,3}$,
  \textbf{Carlos Oliver}$^{4}$,\\
  \textbf{Véronique Stoven}$^{1,2,3}$,
  \textbf{Vincent Mallet}$^{1,2,3,\dagger}$
  \vspace{0.5cm}\\
  $^1$Mines Paris, PSL Research University, CBIO, Paris, France\\
  $^2$Institut Curie, PSL Research University, Paris, France\\
  $^3$INSERM, U1331, Paris, France\\
  $^4$Vanderbilt University, Nashville, Tennessee, US\vspace{0.3cm}\\
\normalsize{$^\dagger$ correspondence to \texttt{vincent.mallet@minesparis.psl.eu}
}}
\date{}

\begin{document}

\maketitle


\begin{abstract}
    Targeting RNA with small molecules offers significant therapeutic potential.
    Machine learning could substantially accelerate preclinical drug discovery, from hit identification to lead optimization.
    Yet a fundamental limitation emerges: drug design machine learning models, tailored for proteins, are not readily applicable to RNAs because of fundamental differences between RNAs and proteins in both structural characteristics and interactions with small molecules.
    
    RNA-specific approaches have consequently emerged, primarily focusing on binding site identification and virtual screening.
    In this review, we comprehensively compare machine learning tools for RNA-targeting drug design according to the tasks they address, their methodology and their relevance in RNA-specific contexts.
    As open challenges will catalyze new method development, we emphasize the need for standardized, drug design-specific evaluation approaches.
    We provide clear guidelines to establish these standards along with a benchmark assessing the ability of current machine learning models to predict specific drug-RNA interactions.
\end{abstract}

\section{Introduction}
\label{sec:intro}
RNA has emerged as a compelling target for small molecule therapeutics \cite{childs2022targeting, thomas2008targeting}. Beyond the canonical role of messenger RNAs in genetic information transfer, non-coding RNAs play pivotal roles in gene expression regulation \cite{matsui2017non} and are involved in numerous human disease pathways \cite{toden2021non}. RNAs have proven to be targetable by small molecules \cite{mccown2017riboswitch, childs2016approaches}, paving the way for novel therapeutic avenues, particularly against diseases involving undruggable proteins \cite{warner2018principles}. Despite this potential, RNAs remain significantly under-exploited as therapeutic targets. A few antibiotic families targeting ribosomal RNAs (rRNAs) successfully reached the market because these RNAs are well structured and because the prokaryote ribosomes differ substantially from those of eukaryotes \cite{cai2025discovery}. However, the first FDA-approved drug targeting a non-ribosomal RNA (risdiplam, a pre-mRNA splicing modifier against spinal muscular atrophy) was only authorized in 2020, and remains the only one to date.

At the same time, drug design-focused machine learning models are emerging. In virtual screening especially, their computational efficiency allows for the exploration of vaster chemical libraries, significantly increasing the probability of identifying high-affinity compounds. This holds significant potential for accelerating RNA-targeting drug design.

However, the field has been historically dominated by protein targets, leading to the development of highly successful machine learning methods that are fundamentally protein-centric (e.g., for binding affinity prediction \cite{pignet, interactiongraphnet, deepdta, graphdta} and computational docking \cite{equibind, diffdock, surfdock}). These approaches cannot be directly applied to RNA, which presents fundamental differences that necessitate specialized modeling strategies.

\subsubsection*{On the uniqueness of RNA}
RNAs exhibit greater conformational flexibility than proteins, complicating their experimental structure determination through X-ray crystallography \cite{afm}. These experimental challenges contribute to the scarcity of RNA structural data: as of 2025, the Protein Data Bank (PDB) \cite{pdb} contains 232,962 protein structures compared to only 8,767 RNA structures. Moreover, RNA-small molecule interactions are governed by distinct physicochemical principles. In particular, they are dominated by electrostatic interactions and $\pi$-stacking, whereas protein-small molecule interactions rely more heavily on hydrophobic contacts \cite{kovachka2024small}. Besides, evidence suggests that specific RNA binding sites are more polar \cite{donlic2022r} and more deeply buried within their structures \cite{weeks} than protein binding sites.

\begin{figure}[htbp]
    \centering
    \includegraphics[width=\linewidth]{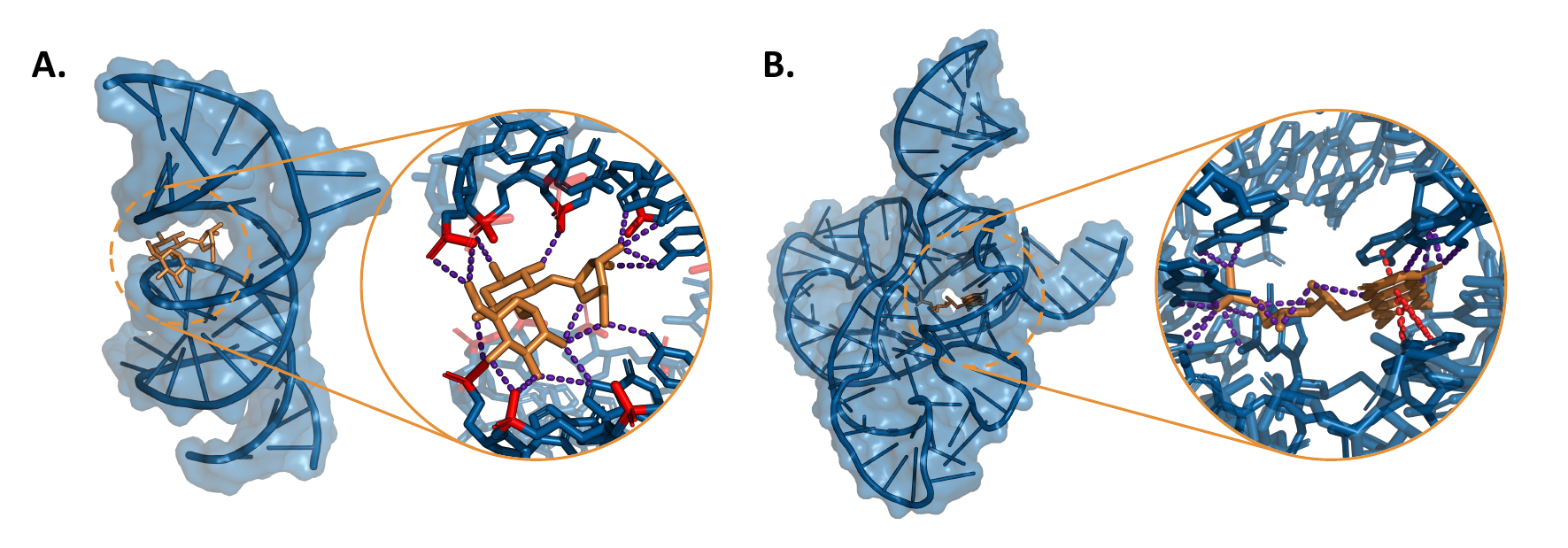}
    \caption{
    Structures of nonspecific (\textbf{A}) and specific (\textbf{B}) RNA-small molecule binding sites. Purple dashed lines represent H bonds and red dashed lines represent $\pi$-stacking. Figures were generated using PyMol \cite{pymol} and fingeRNAt \cite{fingernat}. \textbf{A.} Structure of the 16S-RRNA A-Site unspecifically bound to Ribostamycin (PDB 2et5). \textbf{B.} Flavin mononucleotide (FMN) riboswitch bound to FMN (PDB 2yie).
  }
    \label{fig:binding_site}
\end{figure}

Moreover, several RNA properties favor non-specific interactions: the limited chemical diversity of RNA monomers compared to that of amino acids in proteins \cite{falese2021targeting}, the propensity of RNA to form electrostatic interactions because of the high electronegativity of the backbone \cite{kovachka2024small, padroni2020systematic}, and the importance of $\pi$-stacking \cite{warner2018principles}. These characteristics make specificity a critical consideration in RNA-targeting drug design. Non-specific RNA binders include intercalators, interacting with RNAs through $\pi$-stacking and hydrophobic contacts, and aminoglycosides. For instance, Figure \ref{fig:binding_site}A illustrates a non-specific RNA-aminoglycoside interaction (between ribostamycin and the 16S ribosomal RNA A-Site), that relies on many H-bonds involving atoms from the phosphate groups of the RNA (colored in red in the figure), which are not sequence-specific \cite{nonspecific_site}.

However, some RNAs display specific binding to small molecules, especially through hydrogen bonds involving RNA bases. For instance, the Flavin mononucleotide (FMN) riboswitch, which binds specifically to the FMN molecule, displays a highly buried binding site (Figure \ref{fig:binding_site}B) \cite{warner2018principles}. The ligand engages with nucleotides across five different loops. Moreover, these interactions involve the Watson-Crick faces of some nucleobases, which are known to form highly specific bonds \cite{specificity}. These complex contacts spanning multiple loops result in a very specific RNA-ligand interaction.

Given these fundamental differences between RNA and protein targets, the development of RNA-specific machine learning approaches is essential. The rapid emergence of novel methods across diverse tasks in the drug discovery pipeline, employing varied modeling approaches and datasets, necessitates a systematic organization of the field based both on the task and on the methodological approach.

\subsubsection*{Preclinical drug discovery pipeline}

The preclinical drug discovery pipeline comprises several steps (Figure \ref{fig:pipeline}A). First, a target (in our case, an RNA involved in the pathology to be treated) must be identified and validated (\textit{Target identification and validation}).

Once a target has been validated, \textit{Hit identification} focuses on identifying promising small molecules (hits) that show demonstrated affinity to this target \cite{hughes2011principles}. A common approach to hit identification relies on screening campaigns, which assess the ability of molecules from a chemical library to bind to the target. However, the drug-like chemical space is so vast (between $10^{20}$ and $10^{24}$ molecules up to 30 atoms \cite{chemical_space}) that it cannot be experimentally tested, calling for efficient in-silico methods.

The pipeline continues through two key refinement stages. During the \textit{Hit-to-lead phase}, hits are chemically refined to significantly increase their affinity and selectivity, generally reaching nanomolar affinity; the resulting compounds are called lead compounds \cite{hughes2011principles}. During \textit{Lead optimization}, lead compound structures are further refined to optimize their pharmacological profile, including properties like synthesizability, delivery, and reduced side-effects, while preserving high affinity and selectivity. Successful compounds become preclinical drug candidates, ready to enter the translational investigation stage.

\subsection*{Overview}

This review systematically surveys machine learning methods for RNA-targeting drug design, focusing specifically on the stages spanning from hit identification to hit-to-lead optimization. Our discussion is organized around four key questions:

\begin{itemize}
    \item \emph{What are machine learning contributions to RNA drug design?} Section \ref{sec:tasks} comprehensively categorizes existing methods by the specific drug design problem they address and identifies research gaps. 
    \item \emph{What are the possible machine learning methods tailored for RNA data?} We examine the possible mathematical representations and machine learning encodings of RNAs that can be used to solve the above tasks, and use this perspective to dissect existing literature in Section \ref{sec:encodings}. 
    \item \emph{What data are available and should be used to train our models?} We discuss best practices for data collection and partitioning, along with existing benchmarking datasets in Section \ref{sec:datasets}.
    \item \emph{How do we rigorously assess model performance?} Our final Section \ref{sec:evaluation} reviews model evaluation in drug design contexts and highlights crucial pitfalls. We notably investigate a critical yet underexplored question in RNA binding affinity prediction: whether current models genuinely capture specific RNA-small molecule interactions or instead rely primarily on ligand features. We propose a new performance assessment and use it to compare four state-of-the-art models.
\end{itemize}

Our analysis serves a dual purpose: to guide researchers in RNA drug design toward the most suitable computational approaches for their work and to offer the machine learning community insights into design principles and research gaps in machine learning for RNA-targeting drug design.

\section{Drug design problems addressed}
\label{sec:tasks}

Machine learning can contribute in several ways to drug design throughout the preclinical drug discovery pipeline.

\begin{figure}[bht]
\centering
\includegraphics[width=\linewidth]{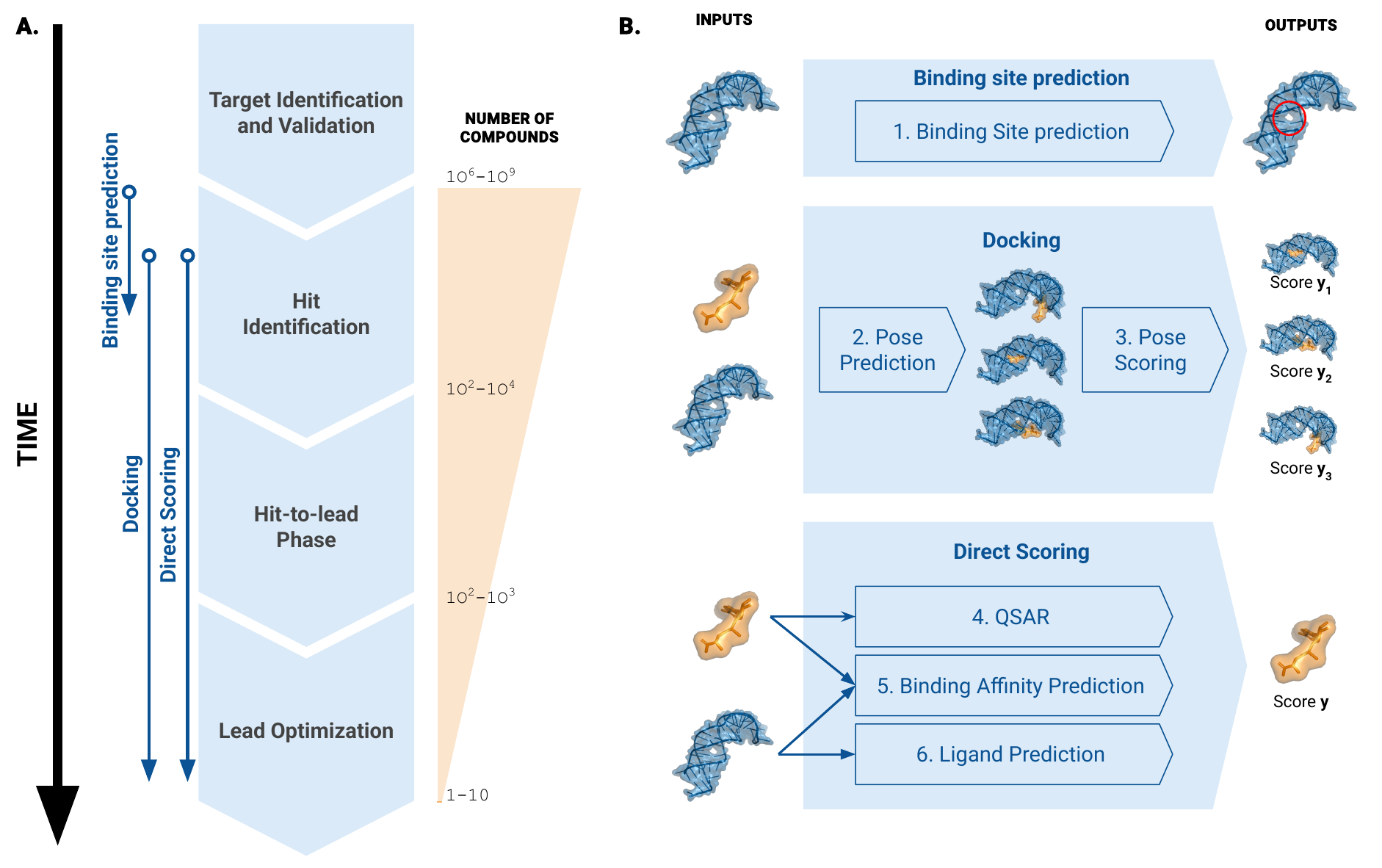} 
\caption{Overview of the preclinical drug discovery pipeline (\textbf{A}) along with six machine learning tasks (\textbf{B}) assisting it. Binding site prediction (1) enables researchers to focus on subparts of an RNA target. Docking, resulting from the joint use of the pose generation (2) and pose scoring (3) tasks, outputs the bound structure of an RNA-ligand pair along with a binding propensity score. Direct scoring methods (QSAR (4), Binding Affinity Prediction (5) and Ligand Prediction (6)) provide a binding propensity score without predicting the structure. They differ by the input data they use.}
\label{fig:pipeline}
\end{figure}

In target identification, machine learning can help to elucidate RNA functions and regulatory pathways \cite{rnagenesis, resm}. Once targets have been identified, machine learning models predict the location of potential binding sites on the target. In hit-to-lead and lead optimization, models predict the structure of the RNA-ligand complex (RNA-ligand pose), thus facilitating rational and structure-based drug design: the knowledge of the pose enables chemists to identify possible and suitable structural refinements \cite{childs2022targeting}. 

Moreover, machine learning's key contribution to preclinical drug discovery is small molecule scoring, which consists in attributing a score to a small molecule assessing its properties. This review focuses specifically on models scoring small molecules based on their predicted binding propensity. Binding scores are relevant to several stages of the drug discovery pipeline. Indeed, during hit identification, small molecule scoring allows researchers to perform virtual screening: they rank compounds based on their scores, thus focusing further experimental assays on the best ranked candidates. During hit-to-lead and lead optimization, binding scores enable chemists to anticipate the impact of molecular optimization on binding. We note that machine learning models have been developed to score small molecules based on other properties such as delivery and side-effects, and refer the reader to specialized reviews \cite{schneider2018automating, vamathevan2019applications}. 

\subsubsection*{Small molecule scoring formulation}
In the present paper, binding propensity is evaluated according to a score denoted $y$. Depending on the application, $y$ can be either a binary indicator of binding or a continuous variable. A binary indicator allows for the exploitation of non-quantitative data, originating from wet-lab assays such as small molecule micro-arrays \cite{uttamchandani2005small}, whereas a continuous variable accounts for the continuous nature of binding energy. 

\medskip

When small molecule scoring is performed during the hit identification stage, it aims to discriminate between active compounds and decoys (presumed non-binders) to pre-filter the chemical library. Therefore, a binary score $y$ is most often used. Conversely, when used during hit-to-lead and lead optimization stages, small molecule scoring must guide fine structural modifications, for which even subtle binding affinity differences become important, therefore, a continuous score $y$ is preferred. The most common choice is the binding affinity, defined as the negative logarithm ($pK_d$) of the dissociation constant ($K_d$) of the RNA-ligand complex.

\subsubsection*{Which drug design tasks can machine learning perform ?}

Machine learning methods for RNA-targeting drug design can be categorized into three families, themselves broken down into six tasks (Figure \ref{fig:pipeline}B).

\medskip

The first family consists of a single task: \textit{binding site prediction} (1), which involves identifying potential binding sites on RNA targets. This step typically occurs during hit identification, helping researchers focus on specific binding sites rather than the whole target during later stages (small molecule scoring and pose generation), thus facilitating these problems.

The second family consists of \textit{docking approaches}, which focus on predicting the target-ligand pose. This family includes \textit{pose generation} (2) (predicting RNA-ligand 3D poses) and \textit{pose scoring} (3) (predicting binding propensity from a pose).

Alternatively, RNA-small molecule binding propensity can be directly estimated, without inferring the pose: we call such methods \textit{direct scoring} methods, which constitute the third family. Direct scoring can be formulated as three distinct tasks: \textit{Quantitative structure-activity relationship} (QSAR) (4) that only takes a ligand as input, \textit{ligand prediction} (5) that only takes an RNA target as input, and \textit{binding affinity prediction} (6), that takes both a ligand and an RNA target as input. 

\medskip

Figure \ref{fig:tasks_representations} and Tables in Appendix \ref{supp:extensive_classification} provide a comprehensive mapping from these tasks to existing computational methods. This mapping helps identify the most relevant approaches for a specific drug design problem.

\subsection{Binding site prediction}
\label{binding_site_prediction}

Prediction of potential ligand binding sites on the target macromolecule is particularly important in structure-based drug design, as it allows subsequent drug discovery efforts to focus on specific binding site structures, thereby improving their efficiency. This problem can be formulated as a machine learning task where the input is an RNA $R$ and the output is a nucleotide-level binary annotation indicating whether each nucleotide participates in a binding site.

Early computational approaches for RNA-small molecule binding site prediction, including Rsite \cite{rsite}, Rsite2 \cite{rsite2} and RBind \cite{rbind}, relied on hard-coded statistical thresholds. 

\medskip

More recently, machine learning methods have been developed to identify RNA binding sites. The current best performers are structure-based geometric deep learning approaches, MultimodRLBP \cite{multimodrlbp} (reaching the highest Matthew's correlation coefficient) and RLBSIF \cite{rlbsif} (highest AuROC). BiteNet \cite{bitenet} also shows interesting performance but on a different dataset. Interestingly, Moller \textit{et al.} \cite{moller2022} proposed a transfer learning approach, leveraging protein examples to identify RNA binding sites. 

SMARTBind \cite{smartbind} is the first sequence-based language model to perform binding site prediction. This avenue holds potential, as it could enable the sequence-based co-folding models to focus on subsets of RNAs, even for RNAs without experimental structure.

\medskip

However, current binding site prediction approaches face critical limitations. Indeed, they are generally trained on \textit{holo} (ligand-bound) RNA structures, whereas experimental researchers often only have \textit{apo} (unbound) structures at their disposal, from which binding sites are harder to infer. RNA conformational flexibility makes the development of \textit{apo}-based models more challenging than for proteins. Molearn \cite{molearn} is a variational autoencoder which generates \textit{holo} conformations of RNAs from \textit{apo} conformations of the same RNAs. Such models could be a stepping stone for \textit{apo}-based models, and using them to build a dataset containing multiple conformations of the same RNAs also represent an interesting direction.

While traditional target-centric methods, such as molecular docking, focus on the binding site, \textit{blind docking} and \textit{co-folding} approaches simultaneously predict RNA-ligand poses and binding sites from complete RNA structures. Encapsulating binding site prediction however annuls the computational advantage of focusing on a substructure.

\subsection{Docking}

\subsubsection{Pose generation}
The \textit{pose generation} task takes an RNA (or RNA pocket) $R$ and a ligand $L$ as input and outputs the 3D structure of the complex $C$ formed by $R$ and $L$. The pose generation task can be formulated in two contexts: known-pocket docking ($R$ is a pocket), where the binding site is predefined and only local search for poses is required, and blind docking ($R$ is a whole RNA), where binding site identification and pose generation must be performed simultaneously.

\medskip

Existing docking software such as DOCK \cite{dock}, AutoDock \cite{autodock} and AutoDock Vina \cite{vina}, were primarily designed for protein-ligand pose generation and rely on physical force fields optimized for protein structures. Adaptations to RNA systems include modified versions of established tools such as DOCKing \cite{docking_rna}, AutoDock \cite{autodock_rna}, and ICM \cite{icm_rna}. Additionally, RNA-specific docking tools have been developed, including MORDOR \cite{mordor}, rDock \cite{rdock}, RLDock \cite{rldock}, and RLDOCKScore \cite{rldockscore}.

A limitation of traditional docking is the computational cost of sampling poses during the pose generation step. To limit this cost, a widely used simplifying assumption is to consider the receptor to be rigid, which is a particularly strong assumption in the case of RNA. Traditional force fields were also shown to have unstable performance and limited accuracy for pose scoring \cite{bender2021practical}. Machine learning methods can be used to address these limitations and enhance the docking pipeline, primarily by performing pose generation through two distinct approaches: \textit{deep docking}, relying on a structure-based RNA representation, and \textit{co-folding}, relying on a sequence-based RNA representation as input.

\medskip

Deep docking aims to predict the biomolecule-ligand pose taking as input separate 3D structure-based representations of the biomolecule (or a pocket of it) and of the ligand. It can either be framed as a regression task, in which a unique pose is being predicted from a ligand-target couple, or as a generative task, where a pose generative model is trained conditioned on a ligand-target couple. Whereas several deep docking models have been proposed for protein-small molecule docking, including EquiBind \cite{equibind}, DiffDock \cite{diffdock} and FlexDock \cite{flexdock}, no such model currently exists for RNA-small molecule docking.

\medskip

Co-folding aims at predicting the RNA-ligand pose from RNA sequence and ligand Simplified Molecular Input Line Entry System (SMILES) \cite{smiles} representations (encoding the 2D chemical structure of the ligand). Therefore, despite requiring 3D structural supervision during training, a co-folding model enables prediction of RNA-small molecule poses without requiring prior knowledge of their 3D structure. It can also potentially capture the specific conformational changes induced upon ligand binding, which is valuable in RNA-targeting drug design given the flexible nature of RNAs.

This has led to the development of several recent co-folding models, trained on diverse biomolecules (proteins, RNAs, DNAs) and their complexes, including biomolecule-biomolecule and biomolecule-ligand interactions. RoseTTAFold All-Atom \cite{rosettafoldAA}, AlphaFold3 \cite{alphafold3}, Chai-1 \cite{chai1}, and Boltz-1 \cite{boltz1} can all perform RNA-small molecule co-folding using generative approaches such as diffusion models. 

Despite these advances, the inherent flexibility of RNAs that makes co-folding valuable also complicates the prediction task. Indeed, RNAs can adopt multiple energetically favorable conformations \cite{story_folding}, making it challenging to predict the actual 3D structure of the RNA-ligand complex under consideration. Moreover, co-folding remains computationally expensive, ranging from 5.9 GPU minutes to 1.5 GPU hours for models like AlphaFold3 \cite{alphafold3}, though some, like Boltz-2 \cite{boltz2}, achieve faster prediction times (20 GPU seconds). Additionally, most models require multiple sequence alignments (MSAs) of RNAs as inputs, which cannot be computed for all RNAs, especially orphan RNAs, and are computationally expensive to generate. Finally, the performance of co-folding models critically depends on the quality of MSA inputs \cite{liu2025boosting}.

\subsubsection{Pose scoring}
\label{pose_scoring}
In \textit{pose scoring}, a machine learning model is trained to predict binding propensity from docking poses. This task takes an RNA-ligand complex $C$ as input and outputs a score $y$ assessing binding propensity. The trained model can then be used as a learned scoring function that replaces the scores based on physical force fields used in classical docking approaches.

\medskip

Early statistics-based RNA-small molecule scoring functions, including LigandRNA \cite{ligandrna}, $\allowbreak$ DrugScoreRNA \cite{drugscorerna} and SPA-LN \cite{spaln}, relied on statistical potentials fitted to experimentally determined poses using inverse Boltzmann relation. These were followed by machine learning approaches, primarily relying on traditional machine learning models using interaction fingerprints \cite{siftml,rnaposers,annapurna,ssmd,rmsdxna}. The only structure-based machine learning approach to RNA-small molecule pose scoring was proposed by RLaffinity \cite{rlaffinity}.

The combination of established physics-based docking programs with machine learning-based scoring functions offers the potential to merge docking accuracy with machine learning's ability to capture complex dependencies that exceeds previous scoring functions. However, this approach remains computationally expensive and time-consuming as the pose generation requires approximately one minute per compound using docking.

\medskip

Co-folding models can be trained jointly with a pose scoring model, showing promising capabilities in virtual screening. AlphaRank \cite{alpharank}, by fine-tuning the co-folding model Protenix \cite{protenix} for protein binding, outperformed specialized models on the JACS 8 \cite{jacs} lead optimization dataset in a continuous binding affinity prediction setting. Boltz-2 \cite{boltz2} inherently includes binding affinity prediction in addition to co-folding and also outperforms specialized models, both in a virtual screening setting on a subset of the MF-PCBA dataset \cite{mfpcba} and in a hit-to-lead and lead optimization setting on a subset of the FEP+ benchmark \cite{ross2023maximal}. However, its performance has not been assessed on RNA-ligand benchmarks. Therefore, we propose a first evaluation of its performance in RNA-ligand binding affinity prediction in Section \ref{ablation_study}.

\subsection{Direct scoring}
Small molecule scoring can also be performed directly, without inferring the pose as an intermediary step: this consists in direct scoring, which can itself be formulated as three distinct tasks given the nature of its inputs.

\subsubsection{Quantitative structure-activity relationship (QSAR)}
The direct scoring of small molecules can be implemented through \textit{quantitative structure-activity relationship} (QSAR) modeling, a widely used method in practice \cite{qsar_review}. QSAR formulates direct scoring as a machine learning task where the only input is a small molecule $L$. It is a purely ligand-based optimization technique that does not require access to RNA structures or features. It abstracts the mechanistic understanding of RNA-ligand interactions, but relies on prior knowledge about ligands for the RNA of interest.

Depending on the ligand training database, this approach can be used to predict the ability of a compound to bind RNAs in general \cite{yazdani2023}, within specific RNA families \cite{rizvi2020targeting, schemnet}, or for individual RNAs \cite{grimberg2022, qsar, haga2023}.

\medskip

Nevertheless, QSAR displays limitations. Abstracting from the drug-target interaction complicates scaffold hopping and activity cliffs modeling \cite{activity_cliffs}. Thanks to their ability to capture nonlinear relationships, QSAR models relying on neural networks with expert-based \cite{schemnet, grimberg2022} or structural \cite{haga2023} features as input, might be more robust than linear QSAR models \cite{qsar} in these difficult but important cases.

Moreover, training QSAR models requires prior knowledge of many binding affinities of small molecules for the considered RNA. This makes such models valuable in late stages of hit identification or for the hit-to-lead and lead optimization phases, when experimental affinity data have already been collected for the target RNA, but limits their applicability at early stages of hit identification, when only a small number of known binders are available for training. In this context, incorporating explicit information about the target, thus allowing transfer of knowledge acquired on various RNA targets to the target of interest, is essential.

\subsubsection{Binding affinity prediction}
\label{binding_affinity_prediction}
To this end, \textit{binding affinity prediction} methods incorporate both an RNA $R$ and a small molecule $L$ as independent inputs (contrary to pose scoring methods) and output a binding score $y$.

The most accurate computational approaches to binding affinity prediction are free energy perturbations. These 3D methods leverage molecular dynamics simulations to compute free energy differences between bound and unbound states, thereby determining RNA-ligand binding affinity. However, these approaches have a high computational cost.

\medskip

Machine learning emerges as an alternative to accelerate binding affinity prediction. Early machine learning models took as input expert-based RNA features based on sequence \cite{smtrs}, secondary structure \cite{smtrs, smajl} or function \cite{rfsmma}. These were specifically designed for one RNA family: microRNAs (miRNAs). Following the rise of structure-based geometric deep learning models for protein-ligand binding affinity prediction \cite{pignet, interactiongraphnet, holoprot}, similar methods appeared for RNA-ligand binding affinity prediction, such as RNAmigos 2 \cite{rnamigos2} and GerNA-Bind \cite{gernabind}, now among the best performing methods.

However, Volkov \textit{et al.} (2022) \cite{volkov} identified a critical limitation of protein-small molecule binding affinity prediction deep learning models: these models tend to learn binding affinity from the small molecule alone (i.e. they tend to behave as QSAR models), rather than learning about protein-small molecule complementarity. Thus, they might favor non-specific over specific binders. Therefore, in subsection \ref{sec:ablation}, we experimentally assess and quantify the tendency of RNA-small molecule binding affinity prediction models to learn from small molecules alone.

\subsubsection{Ligand prediction}
Direct scoring can alternatively be formulated as a \textit{ligand prediction} task, which takes an RNA $R$ as input and outputs a small molecule $L$ likely to bind the RNA $R$.

Early ligand prediction approaches include InfoRNA \cite{inforna} and RNALigands \cite{rnaligands}, which mine RNA motifs and query databases of known RNA secondary structure motif-small molecule interactions. These were followed by deep learning approaches: RNAmigos \cite{rnamigos} and E3NN \cite{E3NN}, which show similar performance on ligand MACCS fingerprints reconstruction. 

\medskip

As well as binding affinity prediction approaches, ligand prediction approaches can be applied to targets in the absence of known binders. However, ligand prediction models typically generate only one ideal ligand per target. This limitation impedes comprehensive exploration of chemical space and fails to capture the diversity of candidate binding modes and ligands for a given target. To address this limitation, conditioning small-molecule generative models on structural information about the target represents a promising direction that could better account for the multiplicity and variability inherent in ligand-target interactions. While this approach has gained significant attention in protein research \cite{3dgraphsbdd, targetdiff, diffsbdd, difflinker}, it remains unexplored for RNA targets.

\section{RNA encodings}
\label{sec:encodings}

We defined six machine learning tasks supporting drug discovery, as presented in Figure \ref{fig:pipeline}. Each of these tasks (except QSAR) involves RNAs as inputs or outputs. However, biological objects such as RNAs are not inherently machine-understandable. Therefore, an \textit{encoding} is needed: the mapping of biological objects to machine-readable vectors.
The choice of the encoding represents a core part of machine learning model design, since its relevance to the specific task is critical for capturing meaningful information from the data and achieving high prediction performance.
Encoding involves two steps. First, we map each object to a mathematical \textit{representation} (e.g., a scalar, a vector, a graph, or text). Second, we apply a relevant machine learning model (referred to as an \textit{encoder}) to this representation. For instance, if the selected representation is the nucleotide sequence, the encoder must be chosen among models tailored for sequence or text data, such as a Transformer or a 1D Convolutional neural network (CNN). The encoder determines the types of mappings that can be learned between the chosen representation and the final vector.

In the context of RNA-targeting drug design, the biological objects to encode are small molecules, RNAs, and RNA-small molecule complexes. The encoding of small molecules has been extensively studied in chemoinformatics, including in machine learning contexts for drug discovery, with recent reviews available \cite{wigh2022review, chen2023artificial}. Since these molecular representations are readily applicable to RNA drug design, we focus in this section on RNA encodings.

\medskip

The multi-level nature of RNA structure (primary, secondary, tertiary, quaternary) offers a rich landscape of mathematical modeling possibilities. We first detail RNA sequence-based mathematical representations and their associated machine learning models, then we present RNA structure-based representations and their associated machine learning models.

\begin{figure}[htbp]
\includegraphics[width=\linewidth]{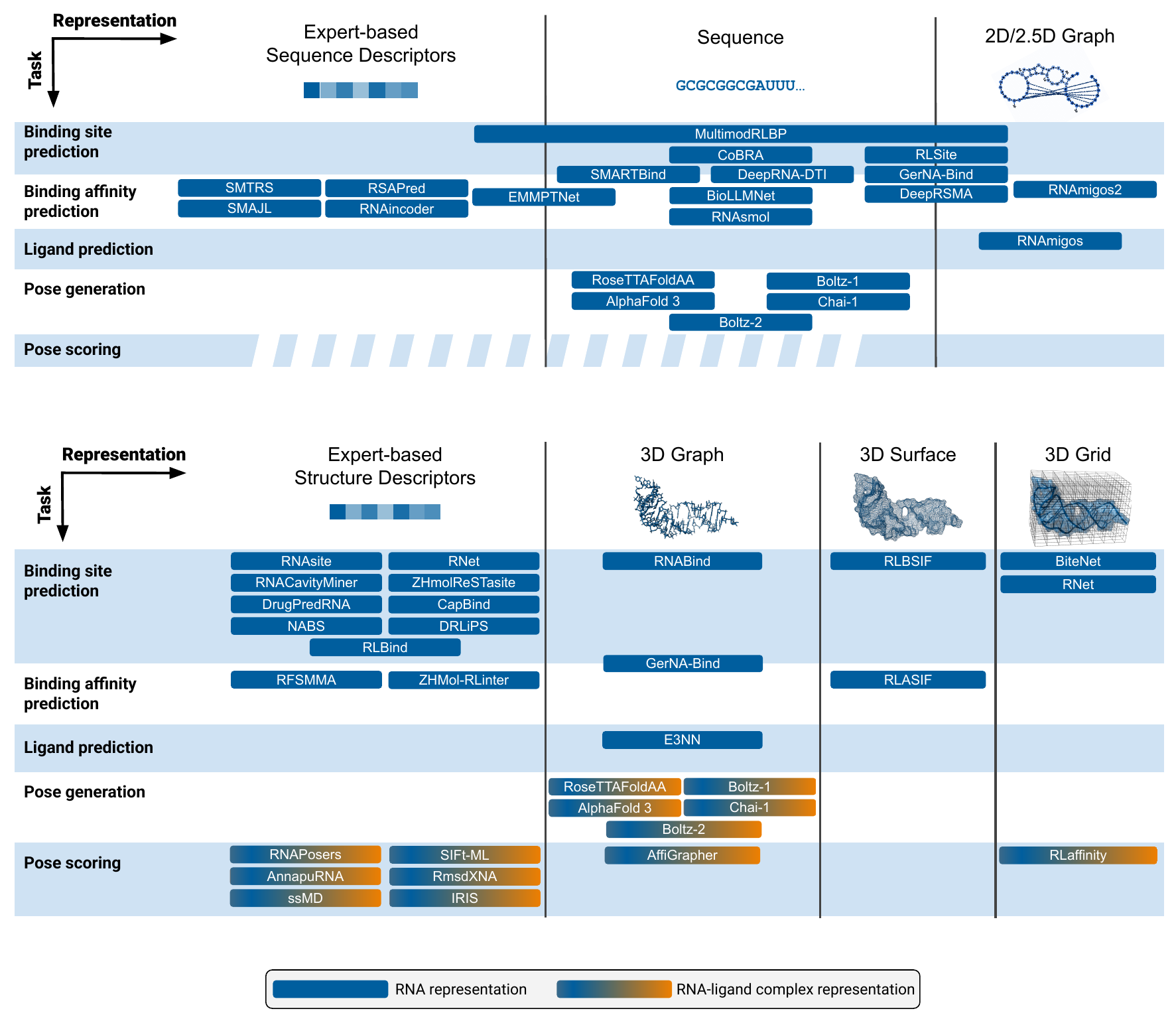} 
\caption{
    Overview of machine learning methods for RNA-targeting drug design. Methods are categorized according to the RNA and complex mathematical representations they rely on (columns). The upper table represents methods relying on sequence-based RNA representations and the lower one represents methods relying on structure-based RNA representations. RNA representations are depicted by blue boxes and RNA-ligand complex representations are represented as blue/orange boxes. Methods are also placed according to the drug discovery task they perform (rows), from hit identification to lead optimization. Note that the QSAR task is not represented in this figure since it does not involve an RNA representation.
}
\label{fig:tasks_representations}
\end{figure}

Figure \ref{fig:tasks_representations} provides an extensive mapping of different machine learning methods based on the representations they rely on. We note that some representations are not compatible with all tasks; for instance, sequence representation is not applicable to pose scoring, which requires input information regarding the 3D pose. Table \ref{supp:RNA_encodings} in the Appendix provides a comprehensive overview of all reviewed methods and their encoding choices for RNAs, small molecules, and RNA-small molecule complexes.

\subsection{Sequence-based encodings}
Since RNAs are uniquely characterized by their sequence of nucleotides, they can be represented by their sequence, also called primary structure.

\subsubsection{Expert-based descriptors}
\label{sec:representations:seq:expert}
The most straightforward approach for encoding RNA sequence relies on a vector of expert-based descriptors. They are often combined with structure-based expert-based features in a single vector, that can be directly used as input to shallow machine learning models such as random forests, support vector machines (SVMs), or linear and logistic regressions. 

These features typically include RNA length and frequency of nucleotide k-mers (substrings of length k of RNA sequences). In addition to k-mers, some methods propose to predict the secondary structure then extract features such as the number of base pairs or secondary structure elements (bulges, hairpins) from this predicted secondary structure. For instance, SMTRS \cite{smtrs} predicts the secondary structure using the ViennaRNA package \cite{viennarna} then computes secondary structure-based features and RSAPred \cite{rsapred} extracts features using RepRNA server \cite{reprna}, which itself relies on secondary structure prediction to compute such features. Sequence-based features might also include features encoding the evolutionary conservation of nucleotides from multiple sequence alignment (MSA). For instance, CapBind \cite{capbind} computes a feature quantifying the evolutionary conservation at each nucleotide position in the RNA sequence using MSA data.

The vast majority of machine learning methods for RNA-targeting drug design rely on expert-based features calculated from the RNA sequence. Such methods draw on solid domain knowledge and are highly interpretable and computationally efficient. However, they represent RNAs with fixed, predefined features that might oversimplify their nature and hence fail to capture complex dependencies that govern RNA-ligand interactions.

\subsubsection{Raw sequence representations}
An alternative approach represents RNA based on the one-letter base code of its raw sequence of nucleotides. This sequence, encoded as textual data, can then be fed to machine learning encoders with architectures adapted from natural language processing. Such machine learning models are expected to learn longer range dependencies than the ones captured by models relying on expert-based features such as k-mer frequencies.

This approach has the advantage of being amenable to systems with no 3D solved structures, and hence to rely on large-scale datasets. This led to a proliferation of RNA language models in recent years \cite{rnabert, rnafm, rnaernie, unirna, evo, evo2}. Such generalist models can encode RNA sequences into embeddings (vector representations), which can subsequently be used for drug design applications. For instance, the specialized models DeepRSMA \cite{deeprsma} and GerNA-Bind \cite{gernabind} rely on the generalist RNA language model RNA-FM \cite{rnafm} to encode RNA sequences.

\subsection{Structure-based encodings}
\label{structure_based_encodings}
Alternatively, several approaches, observing that RNA binding properties greatly depend on structural characteristics \cite{childs2022targeting, weeks}, propose RNA representations that explicitly incorporate structural geometry. In protein research, structure-based approaches outperform state-of-the-art sequence-based models on several machine learning tasks relevant to drug design \cite{gearnet}, providing grounds for optimism regarding similar RNA representations.

\subsubsection{Expert-based descriptors}
As for sequence (Section \ref{sec:representations:seq:expert}), the most straightforward encoding of RNA secondary or tertiary structure is a set of expert-based descriptors, which is most often combined with sequence-based expert features within a single vector.

Structural expert-based features generally include geometric descriptors of nucleotides, for instance the Laplacian norms of nucleotide coordinates in RNASite \cite{rnasite} and ZHMolReSTaSite \cite{zhmolrestasite}. They might also include features describing the size, shape and polarity of accessible surface areas, which are the RNA surfaces accessible to the solvent. For instance, DrugPredRNA \cite{drugpredrna} relies on FreeSASA \cite{freesasa} to compute such features. In addition, some methods propose to construct graphs whose nodes represent RNA nucleotides and edges represent non-covalent interactions and compute features encoding the graph topology. For instance, CapBind \cite{capbind} computes such a graph and uses the degree and closeness of the nucleotide nodes as features.

Some approaches model the RNA-ligand complex rather than RNA alone. In such cases, the feature vectors must capture interaction-specific properties. Relevant complex descriptors include geometric features encoding the relative positioning of RNA and ligand atoms \cite{annapurna, rnaposers, rmsdxna}, docking scores \cite{rnaposers}, and chemical fingerprints \cite{siftml}.

\subsubsection{Full 3D representations}
To tackle complex structural dependencies, alternative approaches represent RNA structure in its full 3D form. Such representations can be combined with geometric deep learning \cite{gdl} (generalization of deep learning to non-Euclidean data, such as 3D shapes and graphs) encoders, thus enabling the modeling of complex relationships. 3D representations are essential for intrinsically geometric tasks such as docking or pose scoring, and can provide valuable geometric information for other drug design applications. These representations encompass various mathematical formulations, including multi-view images, 3D grids, 3D graphs, and surface meshes.

\medskip

Voxel grids generalize 2D images to 3D space, representing structures as sets of voxels (3D pixels). BiteNet \cite{bitenet} adopts this approach for RNA structures, whereas RLaffinity \cite{rlaffinity} applies it to RNA-ligand complexes. Grids enable accurate 3D modeling and leverage well-established convolutional neural networks as encoders. However, this approach is computationally inefficient since the molecular structure occupies only a small fraction of the 3D grid, leading to substantial computational waste.

All-atom 3D graphs model RNA structure as graphs where nodes represent heavy atoms and edges represent covalent bonds. These can be combined with E(3)-equivariant graph neural networks (EGNNs) \cite{EGNN}, which take into account atomic coordinates during learning, as proposed in E3NN \cite{E3NN} or in the Geometric Vector Perceptron (GVP) \cite{gvp}, introduced to learn on protein structures.

Surface meshes represent RNAs as 3D surfaces discretized into triangles, each with distinct features. Building on the biological prior that biomolecular interactions are driven by surface geometry and chemistry rather than internal folds, this representation was initially developed for proteins in MaSIF \cite{masif}, then adapted for RNAs in RLASIF \cite{rlasif} for RNA binding site prediction and RLBSIF \cite{rlbsif} for affinity prediction.

\medskip

However, full 3D representations present specific challenges: high dimensionality and reliance on 3D structural data, which are less abundant than sequence or secondary structure data, can lead to insufficient training data relative to model complexity. RNAs face this challenge even more acutely, as their structures are scarcer than protein structures. Using predicted 2D or 3D structures could overcome this issue. However, this approach introduces a risk of distribution shift relative to experimentally determined structures, as demonstrated for proteins by Huang \textit{et al.} \cite{huang2024protein}. Moreover, current RNA structure prediction performance lags behind state-of-the-art protein structure prediction models \cite{rna_alphafold, stateofthernart, casp16}, thereby limiting the performance of models relying on predicted RNA 3D structures \cite{ablation_representations}.

\subsubsection{Coarse-grained modeling}
Coarse-grained modeling offers an alternative approach by abstracting fine structural details. This introduces a sparsity prior that is likely to reduce the computational complexity of models and tackle data scarcity.

Coarse-grained representations generally take the form of graphs whose nodes represent RNA residues. Edges can link adjacent residues in the backbone and residues involved in a base pair, as proposed in RNAmigos \cite{rnamigos}, or residues closer than a defined radius. Beyond node and edge definitions, RNA graphs can be customized by adding features to the graph nodes and edges or by distinguishing different edge types, which will be taken into account during the learning process.

\medskip

The most straightforward approach uses 2D graphs that encode secondary structure. Their edges represent backbone connections and base pairs. For instance, GerNA-Bind \cite{gernabind} relies on 2D graph modeling, among other representations.

2.5D graphs provide an intermediate level of abstraction: like 2D graphs, their edges encode backbone connections and base pairs, but they also include non canonical edges. Moreover, these graphs possess different edge types accounting for the geometric families of base pairs, as defined by the nomenclature established by Leontis and Westhof \cite{leontis_westhof}. Therefore, this representation incorporates geometric information that is finer than raw base pairing but sparser than atomic coordinates. For example, RNAmigos \cite{rnamigos}, RNAmigos 2 \cite{rnamigos2}, and MultiModRLBP \cite{multimodrlbp} leverage 2.5D graph representations.

Finally, coarse-grained graphs may also be geometric (3D) graphs carrying 3D coordinates as node features. These coordinates could include either residue centroid positions, as proposed in RNABind \cite{rnabind}, or the coordinates of a chosen set of representative atoms. For instance, GerNA-Bind \cite{gernabind} uses 3D coordinates of C4', N1, and P atoms as node features for each residue. In such graphs, edges are generally built based on the distances between nodes (either with a distance threshold or with k nearest neighbors).

\subsection{Multimodal encodings}
Several of these representations can be incorporated within a single multimodal model: a machine learning model that takes different data modalities as input and merges them within its architecture. In protein research, multimodal models such as AtomSurf \cite{atomsurf} or S3F \cite{S3F} proposed to combine sequence, graph and surface representations, outperforming state-of-the-art single-modality approaches.

Several multimodal approaches have been developed for RNA applications. Multimodal models for RNA-small molecule binding affinity prediction include MultimodRLBP \cite{multimodrlbp}, which integrates expert-based features, sequence, and 2.5D graph data; DeepRSMA \cite{deeprsma}, combining sequence and 2D graph representations; and GerNA-Bind \cite{gernabind}, which merges sequence, 2D graph, and 3D graph modalities. Outside drug design, HARMONY \cite{harmony} proposes a general-purpose multimodal model combining sequence, 2D and 3D graph representations.

The integration of embeddings from different modalities can be achieved through various strategies. Simple aggregation functions such as averaging or concatenation provide straightforward solutions, whereas cross-attention mechanisms, employed by DeepRSMA \cite{deeprsma} and RNAsmol \cite{rnasmol}, offer enhanced integration capabilities. Alternatively, some models incorporate RNA sequence language model embeddings as node features within graph representations, as demonstrated in RNAmigos2 \cite{rnamigos2} and RNABind \cite{rnabind}.

\subsection{Comparing RNA representations}
When choosing a representation, several questions should be taken into consideration. Foremost is the representation's expressiveness (i.e., the level of structural and spatial detail it encodes about the RNA). From this perspective, three-dimensional (3D) representations are inherently more expressive than two-dimensional (2D) representations, which in turn surpass sequence-based representations in their level of detail. However, the richness of the representation should be balanced with the data availability. Indeed, the higher the dimension of the representation, the larger the training dataset must be and the more computational power is needed. 

\medskip

RNA structures are far scarcer than RNA sequences (see Section \ref{sec:datasets}). Using predicted 2D or 3D structures offers a solution, but this approach introduces distribution shifts relative to experimentally determined structures, as demonstrated by Huang \textit{et al.} for proteins \cite{huang2024protein}. Moreover, current RNA 3D structure prediction performance lags behind state-of-the-art protein structure prediction models \cite{rna_alphafold, stateofthernart}, subsequently limiting the performance of models relying on predicted RNA 3D structures \cite{ablation_representations}. RNA secondary structure and base pairing predictors \cite{bayespairing,jar3d, viennarna,eternafold} are more accurate than existing 3D structure predictors, making the use of predicted structures a promising avenue for 2D and 2.5D representations specifically.

\medskip

Empirical validation is thus needed to guide representation choices. Two benchmarks of RNA representations have been performed to date. The benchmark by Xu \textit{et al.} \cite{ablation_representations}, relying on predicted structures for structure-based encodings, found 2D graph representations to significantly outperform both sequence-based and 3D representations. Conversely, in our benchmark relying on experimentally determined structures \cite{rnatasks}, the 2D graph representation performed on par with the 3D graph representation, both still outperforming the sequence representation. This suggests that the performance of 3D graph representations on predicted structures was impeded by the inaccuracy of structure predictors. Moreover, we found the 2.5D representation to yield the best results, which might be attributed to the sparsity and biological relevance of this representation.

Beyond existing benchmarks, future work must deepen the experimental comparison of representations. This includes benchmarking across a wider array of tasks, especially those not intrinsically structure-related, and incorporating multimodal representations that combine sequence and structural information. These directions are essential for guiding optimal representation choices.

\section{Datasets and splitting}
\label{sec:datasets}

Apart from the choice of representations and task formulation, the efficiency of machine learning models for RNA drug design relies heavily on the datasets they use for training and testing. Moreover, particular care is required to estimate the performance of learning-based methods, depending on the considered drug discovery task. In this section, we discuss the current best practices and advocate for more rigorous model comparisons.

\subsection{Datasets}

\begin{figure}[htbp]
    \includegraphics[width=\linewidth]{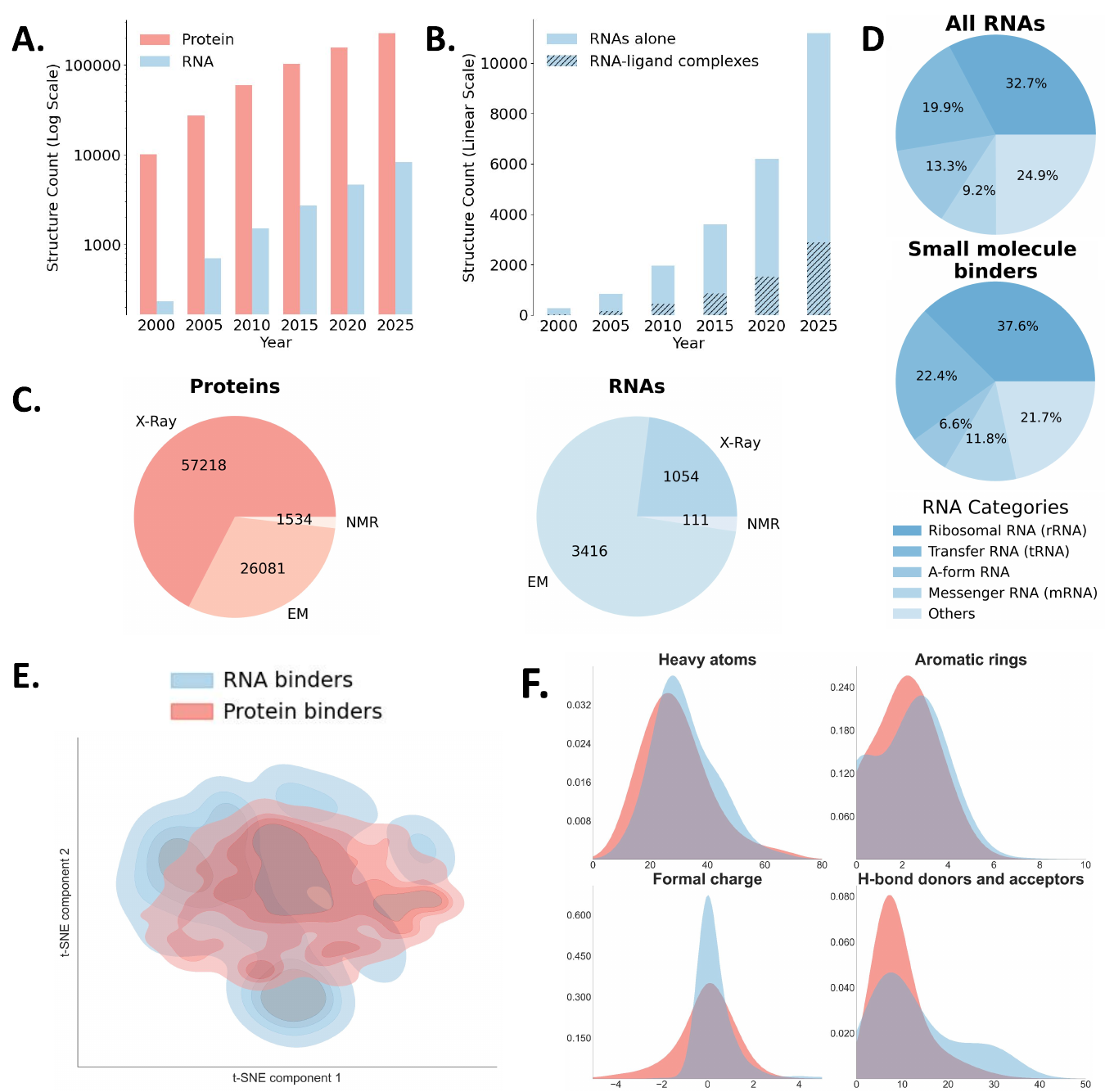} 
    \caption{
        \textbf{A.} Evolution of the number of RNA and protein structures in the PDB.
        \textbf{B.} Evolution of the number of RNA structures in the PDB, both alone and ligand-bound.
        \textbf{C.} Distribution of experimentally determined protein and RNA structures (PDB entries) by primary technique (X-ray, Cryo-EM, NMR), released between 01/01/2020 and 11/27/2025. The specific time window was selected to control for time-dependent biases in the adoption rates of experimental methods.
        \textbf{D.} Repartition of RNA structures available in the PDB by category.
        \textbf{E.} t-SNE visualization of protein binders (in red) and RNA binders (in blue) in the chemical space.
        \textbf{F.} Distribution of chemical properties for proteins and RNA. Upper left: Number of heavy atoms. Upper right: Number of aromatic rings. Bottom left: Formal charge. Bottom right: Number of H-Bond donors and acceptors (total).
    }
    \label{fig:dataset}
\end{figure}

Training and testing machine learning models for RNA-targeting drug design may require access to three types of data: RNA sequence data, RNA structure data, and RNA-small molecule interaction data. In what follows, we detail the existing databases and tools providing such data, along with their strengths and limitations.

RNA sequence data can be retrieved from dedicated sequence databases such as Rfam \cite{rfam} (providing MSAs and family information), RNAcentral \cite{rnacentral} (focused on non-coding RNAs and containing over 35 million sequences), and the NCBI's non-redundant nucleotide sequence database \cite{ncbi}. The largest non-coding RNA sequence database system, RNAcentral \cite{rnacentral}, contains 35,400,760 RNA sequences in its twenty-fourth release (2025).

\medskip

Beyond sequence data, structure information provides another crucial component for RNA analysis. Structural data are mostly extracted from the PDB. They are far scarcer than protein structure data: as of January 1st, 2025, the PDB contained only 8,801 RNA structures, compared to 233,272 protein structures. This stark disparity is illustrated in Figure \ref{fig:dataset}A, which compares RNA and protein structure counts in the PDB from 2000 to 2025. The data scarcity is even more pronounced when excluding ribosomal RNAs, which are highly structured through their interactions with proteins, and therefore easier to determine experimentally (see Figure \ref{fig:dataset}C for a complete breakdown of PDB RNA structures by RNA categories).

The scarcity of RNA structural data stems, in part, from experimental challenges specific to RNA structure determination: because of their higher flexibility, RNAs are harder to crystallize, thus limiting the use of X-ray crystallography. This could explain the smaller share of structures determined by X-Ray among RNA structures compared to protein structures and, conversely, the larger share of structures determined by cryo-electron microscopy (cryo-EM), observed in Figure \ref{fig:dataset}C.

Specific tools facilitate the extraction, preprocessing and use of RNA structures from the PDB. For instance, BGSU \cite{bgsu} removes redundancy among RNA structures, RNAsolo \cite{rnasolo} provides cleaned files in various formats through a web interface and RNANet \cite{rnanet} joins RNA structures with their corresponding MSAs.

\medskip

When using RNA data for drug design, complementary data regarding RNA-small molecule interaction becomes necessary. Restricting to RNA-small molecule complex structures restrains the amount of data available, as illustrated by Figure \ref{fig:dataset}B: as of January 1st, 2025, 3,076 RNA-small molecule complex structures were available, accounting for approximately 35\% of all RNA structures. Several databases provide RNA data specifically curated for drug design applications. Some of them compile RNA-small molecule binding pairs alongside their binding affinities, RNA sequence and expert-based features characterizing small molecules, such as SM2miR \cite{sm2mir} (focusing on miRNAs; 2,925 interactions) and R-SIM \cite{rsim} (2,501 interactions). The protein-focused PDBbind database \cite{pdbbind} also offers 101 RNA-ligand couple data. Other databases provide experimentally determined RNA-small molecule complex structures, such as HARIBOSS \cite{hariboss} (1,000 structures) and ROBIN \cite{yazdani2023}. 

Drug design-specific preprocessing may encompass additional steps, such as removing protein-dominated complexes and selecting drug-like small molecules based on their mass or chemical properties \cite{moller2022}. By comparing all RNA and protein binders contained in PDBbind \cite{pdbbind} and R-SIM \cite{rsim} databases (Figures \ref{fig:dataset} E and F), we observe that they follow different distributions in the chemical space. This underlines the need for machine learning models specific to RNA-targeting drug design, as well as the use of dedicated chemical libraries.

In particular, RNA binders cannot be negatively charged, due to the high negative charge of RNA backbone. They include more hydrogen bond donors and acceptors, as they cannot rely on hydrophobic interactions for binding. They also tend to possess more aromatic rings than protein binders, which highlights the role of $\pi$-stacking in RNA-small molecule interactions. We propose a user-friendly web browser for a more extensive exploration of RNA- and protein-binder chemical spaces on the following website: \url{https://wisskarrou.github.io/RNA_vs_protein_binders/}.

\subsection{Splitting}

\subsubsection{Motivation}
Once a dataset has been selected and preprocessed, a key step for machine learning follows: dataset splitting. Indeed, machine learning consists of three steps: model training (learning of the input-output relationship by the model), validation (first range evaluation of the model aiming at calibrating hyperparameters, which defines the type of input-output relations the model is allowed to learn) and test (final model evaluation aiming at giving the users an idea of the confidence they can expect and comparing its performance with alternative models). If training, validation and test are done on the same dataset, it is likely that the learned model will be specific to this dataset: this phenomenon is called \textit{overfitting}. Therefore, the performance of this model on external datasets will be much lower than the performance reported during the test phase. Since machine learning models are generally meant to be used on external datasets, such as a user's proprietary data or new RNA targets being unveiled in our case, we aim at fostering generalization ability and limiting overfitting by splitting the dataset into independent training, validation and test sets.

\medskip

Splitting is generally achieved by randomly splitting the dataset into three subsets, thus ensuring no data is duplicated between these sets. However, in structural biology and chemistry, this straightforward approach faces unique challenges. Indeed, two distinct RNA or protein structures might still be highly similar. For instance, they can correspond to a single macromolecule experimentally determined under different conditions or to evolutionarily close macromolecules. In the same way, distinct ligands can show great similarity when they belong to the same chemical scaffold. In other words, the independent and identically distributed (IID) data assumption is not always satisfied in structural biology and chemistry \cite{splitting, atas2023approach}.
Therefore, the choice of the splitting strategy is not straightforward: one might either want objects from the training, validation and test sets to be only distinct (weak generalization) or to enforce a dissimilarity level between them (strong generalization).
The choice of the splitting strategy depends on the intended use of the model. Indeed, the performance metrics reported on the test set aim at reflecting the expected model performance in real conditions. Therefore, the domain shift between training and test sets should mirror the shift expected between training set and real-world application data.

\medskip

This shift depends on the stage of the drug design process and task performed. For instance, a binding affinity prediction model does not need the same generalization according to the context it is used in. Indeed, in early virtual screening, we generally have little data regarding the RNA and ligands of interest, therefore strong generalization is required for both RNAs and ligands. In hit-to-lead and lead optimization, data regarding the target RNA have already been collected during virtual screening. Therefore, a weak generalization might be sufficient for RNA, enabling the use of RNA random splitting, while strong generalization is still needed for ligands. Conversely, when repurposing a drug for a distinct target, the model should be able to generalize to new targets but not to new ligands. In these cases, weak generalization might be sufficient for ligands and strong generalization should be preferred for targets \cite{volkov}. Given these application-specific requirements, researchers have developed several splitting strategies. 

\subsubsection{Similarity-aware splits}
Enforcing RNA, protein or ligand strong generalization can be achieved through a similarity-based split: molecules are clustered according to their similarity and molecules belonging to the same cluster only appear in the training, validation or test set. Experiments by Li \textit{et al.} (2017) \cite{splitters} and Yang \textit{et al.} (2020) \cite{yang2020predicting} on proteins and Wyss \textit{et al.} (2025) on RNAs \cite{rnatasks} demonstrated that random splitting leads to significantly inflated performance metrics compared to similarity-based splitting. 

\medskip

RNA similarity-based splitting could be defined either in terms of sequence- or structure-based similarity. For instance, RLBind \cite{rlbind}, CapBind \cite{capbind} and MultiModRLBP \cite{multimodrlbp} group RNAs based on sequence similarity using CD-Hit clustering \cite{cdhit}. Other methods perform structural similarity-based computing by clustering RNAs based on their structural similarity, as measured by the TM-Score \cite{tmscore}, used in RNABind \cite{rnabind}, and the RMscore \cite{rmalign}, used in DRLiPS \cite{drlips}. 

To enforce generalization across ligands, ligand similarity-based splitting clusters ligands using chemical similarity metrics. For instance, EMMPTNet \cite{emmptnet} and AnnapuRNA \cite{annapurna} rely on the Tanimoto index. Another approach to ligand similarity-based splitting lies in scaffold splitting: grouping molecules based on the Murcko scaffold \cite{murcko}.

In order to tune precisely the desired degree of generalization needed according to the expected similarity between training data and real conditions datasets, a recent approach named similarity aware evaluation \cite{splitting} can be leveraged. It proposes to build a test set with a user-defined distribution of molecules within each bin of similarity with the train set. This is particularly valuable in situations where the model is meant to be used on datasets showing both similar and dissimilar molecules to the ones of the training dataset.

\subsubsection{Time split}
As similarity-based split may be too strict and make the task excessively difficult compared to real conditions (especially in settings when the intended use of the model is on RNAs from the same family as those of the training set), an alternative approach to RNA or protein data splitting was proposed: time split \cite{timesplit}.

Time split \cite{timesplit} amounts to splitting RNA or protein structures based on their release date (in the PDB for instance). It aims at replicating how data emerges during drug discovery, where only previously determined structures can predict new ones, and avoiding training and testing on very similar structures which would have been released simultaneously in the PDB as part of a same experiment. Temporal splitting has been popularized by AlphaFold \cite{alphafold} and widely used in protein pose generation \cite{equibind, diffdock} as well as in structure prediction and co-folding \cite{alphafold3, boltz2}. In RNA research, it is used by RNABind \cite{rnabind}.

\subsubsection{Recommendations}
Whereas many RNA-targeting drug design machine learning models still rely on random splitting, we advocate to systematically discuss and ground the splitting strategy choice in machine learning papers for RNA-targeting drug design.

The choice of the splitting strategy should be guided by the desired degree of generalization, which in turn depends on the intended use of the model in real conditions. 

Alternatively, new models should be evaluated and compared with various splittings strategies, to provide performances expected in various real-case situations, as proposed in GerNA-Bind \cite{gernabind}.

\subsection{Benchmarks}
In machine learning for RNA-targeting drug design, as often in machine learning, multiple computational models are developed to tackle the same tasks. Therefore, it is crucial to enable fair comparison between those models: testing them on the same dataset, adopting the same splitting strategy and measuring the same evaluation metrics. We define a \textit{benchmark} as the combination of a dataset, a splitting strategy and a set of evaluation metrics.

\medskip

Unfortunately, the majority of papers rely on their own datasets and splitting strategies, hindering fair comparison across models. This fragmentation significantly impedes the field by hindering the identification of the most promising modeling approaches, slowing overall progress, and potentially leading to redundant research efforts.

This fragmentation contrasts sharply with the more mature protein drug design field, which has successfully established several standard benchmarks. These include PDBbind \cite{pdbbind}, Davis \cite{davis} and KIBA \cite{kiba} for binding affinity prediction, CrossDocked \cite{crossdocked} and PLINDER \cite{plinder} for protein-ligand docking tasks. These standardized benchmarks have enabled systematic progress in protein drug design by allowing researchers to build directly upon previous work and identify genuine advances.

\medskip

Recognizing this critical gap in RNA research, recent efforts have begun developing similar standardized resources. In binding site prediction, the TR60 and TE18 datasets from RNASite \cite{rnasite} have emerged as standards. Moreover, tools such as RNAglib \cite{rnatasks} aim to address this gap by providing standardized datasets, splitting strategies, and evaluation protocols across multiple RNA-related tasks, including binding site and binding affinity prediction. However, comprehensive benchmarks for sequence-based models remain an important area for future development.

\section{Evaluation}
\label{sec:evaluation}

In drug discovery, the adoption of a new model hinges on its convincing performance evaluation. While standard metrics are useful, they must be complemented by domain-specific measures that demonstrate real-world applicability. Ultimately, evaluation metrics should reflect how models will be used in practice to ensure meaningful comparisons and accurately assess their utility. Therefore, the choice of metrics is highly dependent on the task performed.

\subsection{Evaluation metrics}

Binding site prediction is often framed as a node-level classification task, as the binding site is defined by the list of atoms or nucleotides. Therefore, the evaluation generally relies on standard classification metrics, such as Area under Receiver-operating curve (ROC-AUC), precision, or accuracy. Given the significant class imbalance between binding versus non-binding sites, we emphasize metrics that account for imbalance, including Matthews correlation coefficient (MCC), F1-score, and balanced accuracy.

Pose generation tasks (deep docking or co-folding) must be assessed according to their ability to recover a structure close to the experimental structure. Models such as AlphaFold 3 \cite{alphafold3} and Chai-1 \cite{chai1} assess their co-folding performance through ligand RMSD: after aligning the protein or RNA atoms between structures, they compute the RMSD between predicted and experimental ligand pose.

When assessing pose scoring model capabilities, we must assess models' ability to identify native-like conformations among a set of docking-generated poses. Therefore, a commonly used metric is the success rate: the success rate at X Å threshold denotes the percentage of RNA-ligand complexes whose best-scored pose (according to the pose scoring model) has a lower ligand RMSD than X Å to the native pose. This metric is for instance used in AnnapuRNA \cite{annapurna} and RNAPosers\cite{rnaposers}.

Finally, direct scoring methods (binding affinity prediction, ligand prediction or QSAR) can be assessed through standard classification or regression metrics, depending on the binary or continuous nature of their target variable. 

\medskip

Furthermore, these methods can be assessed using virtual screening metrics that better reflect practical applications. To compute such metrics, the models are tested on sets containing true ligands and decoys for each RNA. Then, compounds are ranked by predicted binding affinity. Success is measured by the model's ability to rank native ligands higher than decoys. For instance, E3NN \cite{E3NN}, RNAmigos2 \cite{rnamigos2} and RNAsmol \cite{rnasmol} report the mean normalized rank of native compounds. RNAmigos2 additionally employs the X\%-enrichment factor, denoting the factor by which the fraction of active ligands in the top-scoring X\% of compounds exceeds the overall fraction of active ligands.

Given their closer alignment with practical applications, we advocate prioritizing metrics such as success rates and enrichment factors over purely statistical measures when developing machine learning models for RNA drug design.

To assess the performance of such virtual screening tasks, it is crucial to choose decoys whose distribution in the chemical space reflects the distribution of active compounds. Otherwise, the performance of the model could be overestimated \cite{reau2018decoys}. Many efforts have been developed to select relevant decoys depending on the protein target at hand, for example through the DEKOIS \cite{dekois} or DUD-E \cite{dude} databases, and similar initiatives would be welcome for RNA targets. 

\medskip

Beyond the ability to retrieve a large number of ligands, drug designers are interested in quantifying the diversity of retrieved compounds. Diverse hits provide more opportunities for hit optimization and lower the risk of systematic failure during later stages in the drug design pipeline \cite{zhu2013hit}. The only paper investigating the diversity of its retrieved hits is RNAmigos2, which leverages visual checks and optimal transport to quantify the variety of compounds. We encourage future research to report metrics assessing the diversity of retrieved hits, including the number of unique scaffolds among the retrieved active compounds. For instance, ScaffAug \cite{scaffaug} proposes a metric named $SD_{100}$ that quantifies the scaffold diversity among retrieved hits, along with data augmentation and re-ranking modules which can be plugged in to virtual screening models to enhance the diversity of their hits.

\subsection{Binding affinity prediction ablation study}
\label{sec:ablation}
We performed experiments to fill two gaps in the evaluation of binding affinity prediction models. 
First, the leading co-folding and binding affinity prediction model Boltz-2 \cite{boltz2} has only been assessed on proteins and not on RNAs, whereas it has been trained on RNA-small molecule affinity data. Second, the ability to learn specific interaction patterns rather than learning from ligand information alone is not evaluated by most existing RNA-small molecule binding affinity prediction models. The code and material necessary to reproduce the experiments are available at: \url{https://github.com/wisskarrou/RNA_ligand_ablation_study.git}

\subsubsection{Assessment of Boltz-2 binding affinity prediction capabilities on RNAs}
The technical report for Boltz-2 \cite{boltz2}, a leading co-folding model for predicting binding affinity between proteins or RNAs and small molecules, reports an evaluation of its protein-small molecule prediction capabilities, which are outstanding, with an AuROC greater than 0.80 on the MF-PCBA benchmark \cite{mfpcba}. However, its RNA-small molecule binding affinity prediction capabilities were not assessed. Therefore, we assessed the binary binding prediction performance of Boltz-2 on the ROBIN dataset \cite{yazdani2023} to address this gap. We selected six RNA targets from ROBIN (namely SAM\_ll, ZTP, TPP, PreQ1, NRAS, and RRE2B) and screened all ROBIN small molecules against them, reporting the exact same metrics as those reported for Boltz-2's protein evaluation: the AuROC computed both at the global level (averaged across all RNA-small molecule couples) and target level (average of the per-RNA averages) and the enrichment factors at 0.5\%, 1\%, 2\% and 5\%.

\begin{table}[H]
\centering

\setlength{\tabcolsep}{10pt}
\begin{tabular}{@{}lcccccc@{}}
\toprule
\textbf{} & \multicolumn{2}{c}{\textbf{AuROC}} & \multicolumn{4}{c}{\textbf{Enrichment Factor}} \\
\textbf{} & target & global & 0.5\% & 1\% & 2\% & 5\% \\
\midrule
Boltz-2 & 0.530 & 0.535 & 1.068 & 1.063 & 0.988 & 1.180 \\
\bottomrule
\end{tabular}

\caption{Performance of Boltz-2 on RNA-small molecule interaction prediction, assessed on the ROBIN database.}
\end{table}

Our experiment reveals that Boltz-2 is not performing well in RNA-small molecule binding affinity prediction, with near random metrics, far below the performance it reaches in protein-small molecule binding affinity prediction. This observation underlines the need for RNA-specific approaches towards binding affinity prediction.

\subsubsection{Measuring ligand specificity by target swapping test}

Regarding binding affinity prediction, Volkov \textit{et al.} \cite{volkov} highlighted a tendency of deep learning models to predict binding affinity based solely on small molecule features (see Section \ref{binding_affinity_prediction}). However, standard performance metrics fail to reveal whether models capture meaningful RNA-small molecule interaction patterns or predict affinity from ligand features alone. Therefore, high performance metrics do not preclude the systematic selection of non-specific binders, a particularly pronounced risk for RNAs (see Section \ref{sec:intro}).

Evaluation experiments designed to ensure models learn from macromolecule-ligand interaction patterns rather than ligand features alone, have gained traction in protein research \cite{graphdti, potentialnet, interactiongraphnet, vskrinjar2025have}. In contrast, only RNAmigos 2 \cite{rnamigos2} has performed such validation for RNA-ligand systems. To address this gap, we conducted similar experiments on other open-source RNA-ligand binding affinity prediction models: Boltz-2 \cite{boltz2}, RSAPred \cite{rsapred}, DeepRSMA \cite{deeprsma}, GerNA-Bind \cite{gernabind}, and RNAsmol \cite{rnasmol}.

\medskip

We propose an experimental protocol that systematically swaps RNA-ligand pairings in the test set by applying a permutation $\sigma$ to the RNAs. Each triplet ($R$, $L$, $y$) where $R$ is an RNA, $L$ is a ligand, and $y$ is the experimental binding label (propensity), is replaced by ($\sigma(R)$, $L$, $y$). We then compare model performance on this permuted dataset to performance on the original dataset. 
Unchanged performance despite permuted data suggests that the model relies primarily on ligand features, while decreased performance suggests effective capture of specific RNA-ligand interaction information.
This protocol has the advantage of being applicable without retraining models.
We present the results in Table \ref{ablation_study_quick}. Metrics are reported in two ways: averaged across all RNA-ligand pairs (\textit{global}), and averaged within each RNA target, then averaged across targets (\textit{target}).

A more detailed discussion of the setup and results of these experiments is provided in Section \ref{sec:ablation_study} of the Appendix.

\begin{table}[H]
\centering
\setlength{\arrayrulewidth}{0.3pt}
\begin{tabular}{@{}lccccc@{}}
\toprule
\textbf{} & \multicolumn{2}{c}{\textbf{Metric}} & \textbf{\makecell[t]{Without\\swapping}} & \textbf{\makecell[t]{Target\\swapping}} & \textbf{\makecell[t]{Performance\\gap}} \\
\midrule
Boltz 2 & AuROC & global & 0.535 & $0.533 \pm 0.004$ & -0.4\% \\
& & target & 0.530 & $0.529 \pm 0.002$ & -0.2\% \\
RNAmigos 2 & AuROC & target & 0.899 & $0.791 \pm 0.007$ & -12.0\% \\
RSAPred & AuROC & global & 0.600 & $0.670 \pm 0.057$ & +11.7\% \\
& & target & 0.581 & $0.449 \pm 0.047$ & -22.7\% \\
DeepRSMA & PCC & global & 0.784 & $0.424 \pm 0.064$ & -46.2\% \\
& & target & 0.430 & $0.351 \pm 0.021$ & -22.5\% \\
GerNA-Bind & AuROC & global & 0.783 & $0.514 \pm 0.005$ & -34.4\% \\
& & target & 0.765 & $0.529 \pm 0.005$ & -30.9\% \\
RNAsmol (Mol perturbation) & AuROC & global & 0.974 & $0.975 \pm 0.000$ & +0.1\% \\
& & target & 0.972 & $0.973 \pm 0.002$ & +0.1\% \\
RNAsmol (Net perturbation) & AuROC & global & 0.717 & $0.546 \pm 0.009$ & -23.8\% \\
& & target & 0.651 & $0.569 \pm 0.008$ & -12.6\% \\
\bottomrule
\end{tabular}
\caption{Results of the swapping experiments on RNA-small molecule binding affinity prediction models}
\label{ablation_study_quick}
\end{table}

As can be seen, most models experience a significant performance decrease when the RNA targets are swapped, indicating specificity. This is particularly evident for RNAmigos 2 \cite{rnamigos2}, DeepRSMA \cite{deeprsma}, GerNA-Bind \cite{gernabind}, and RNAsmol with network perturbations \cite{rnasmol}.
However, methods such as RSAPred \cite{rsapred} and Boltz-2 \cite{boltz2} (specifically for RNA-ligand affinity prediction) show limited performance, yielding nearly equivalent results before and after swapping. This suggests a failure to capture specific binding mechanisms.

This outcome emphasizes the necessity to carefully design datasets and decoys. Indeed, when choosing decoys chemically different from RNA binders, the model will likely learn to predict whether a small molecule belongs to the RNA-binding region of the chemical space, overlooking the RNA target. Such models will display low RNA specificity and artificially inflated performance metrics. For future work, we recommend reporting the results of the proposed swapping experiment along with the unperturbed results.

\medskip

To provide a complete analysis, we also performed the symmetric experiment permuting ligands instead of RNAs, resulting in an even stronger performance drop. This indicates that models cannot predict scores from RNA features alone. This could be explained by the data distribution: the test set displays a stable proportion of active compounds per RNA target. Another explanation could lie in the higher expressiveness of ligand encodings, relying on more abundant data and a more mature field than their RNA counterparts, facilitating the direct prediction of the score from the ligand.

\section{Discussion}
\label{sec:outlook}
In this review, we addressed four fundamental questions in machine learning for RNA-targeting drug design: how can machine learning contribute to the drug design pipeline, what are the possible machine learning methods tailored for RNA data, what data should be used and how models should be evaluated.

\medskip

Machine learning can enhance RNA-targeting drug design at several key points of the drug discovery pipeline: hit identification, hit-to-lead phase and lead optimization. Across this pipeline, existing models can detect binding sites, infer RNA-ligand poses (molecular docking), and predict whether they bind each other as well as their binding affinity. However, some widely addressed topics in protein machine learning remain unexplored for RNA, presenting compelling opportunities for further research. Deep docking, widely used in protein research \cite{equibind, diffdock}, could be adapted to RNA structures to predict their 3D bound conformation based on the two individual 3D structures of the RNA and of the ligand. \textit{De novo} ligand design through generative models conditioned on RNA targets or binding pockets could be translated from protein \cite{diffsbdd, targetdiff, pocket2mol, drugflow} to RNA drug design. Such generative approaches offer superior chemical space exploration compared to conventional virtual screening by enabling the creation of novel molecular entities tailored to specific RNA targets. Nevertheless, ensuring synthetic accessibility of computationally generated compounds remains a critical consideration that must be integrated into model development \cite{synthesizability}.

\medskip

To implement these capabilities, machine learning models must rely on machine-understandable representations of RNA. Machine learning approaches to RNA-targeting drug design leveraged the multi-level structure of RNA to propose a wide range of mathematical representations such as raw sequence, 2D structure-based graphs or 3D graphs. However, current machine learning architectures and methodologies largely mirror those used in protein research, suggesting opportunities for RNA-specific innovations. In particular, accounting for RNA conformational flexibility in machine learning models represents a promising research avenue. While flexibility incorporation has already yielded modeling advances in RNA inverse folding \cite{grnade, grnade2} and protein-targeting drug design \cite{lu2024dynamicbind}, it remains unexplored in RNA-targeting drug design. The need for such methodologies will intensify if experimental advances, building upon promising developments such as atomic force microscopy \cite{afm}, ultimately enable high-throughput characterization of RNA conformational ensembles.

\medskip

Beyond methodological considerations, the field faces significant data challenges. Indeed, RNA structures are far scarcer than their protein counterparts. Transfer learning from proteins to RNAs could thus help overcome RNA structural data scarcity. In transfer learning, a machine learning model initially trained on a domain with abundant data (e.g. protein structures) is subsequently specialized (\textit{fine-tuned}) on a second data domain (e.g. RNA structures). Therefore, less data is required in the specialization domain since the model already benefits from the knowledge acquired on the first domain. Pioneering transfer learning approaches were proposed from proteins to RNAs for binding site prediction \cite{moller2022} and across several biomolecules in Atomica \cite{atomica}. Moreover, current models generally rely on specific preprocessing and dataset splitting strategies, hindering fair comparison between models. Therefore, building standardized benchmark datasets for all tasks and ensuring their systematic adoption will be essential to field development.

\medskip

Finally, appropriate evaluation remains crucial for meaningful progress. We advocate the use of application-oriented metrics adapted to the specific stage of the drug discovery pipeline where models are applied. We also emphasize the compelling need to assess and optimize binding affinity prediction models based on their capability to capture significant interaction patterns rather than raw binding affinity values and to retrieve diverse ligands. We hope the ablation study proposed in this review will advance this more nuanced evaluation approach.

\medskip

Together, these considerations outline a clear path forward: leveraging generative modeling for new tasks such as deep docking and \textit{de novo} design, developing RNA-specific methodologies that account for conformational flexibility, establishing standardized evaluation frameworks, and adopting more sophisticated metrics that align with drug discovery objectives. Addressing these challenges will be essential for realizing the full potential of machine learning in RNA-targeting therapeutics.

\section{Acknowledgments}
The authors thank Cyrianne Chabert for her valuable contribution to the ablation study.

V.M. is supported by a Junior Springboard Prairie program, funded by the ANR project ANR-23-IACL-0008. W.K. is supported by Fondation pour la Recherche Médicale (FRM) with the following grant number: ECO202406019160.
This work was performed using HPC resources from GENCI–IDRIS (Grant AD010315435R1).

\printbibliography

\newpage

\appendix

\begin{center} 
\section*{APPENDIX}
\end{center}

\section{Extensive classification of methods}
\label{supp:extensive_classification}
\subsection{Notations}
In this section, we use the following notation: $\mathcal{R}^R$ denotes an RNA \textit{representation}: a function mapping RNAs to mathematical objects. $\mathcal{E}_{\theta}^R$ denotes an RNA \textit{encoder}: a machine learning model with parameters $\theta$ mapping these representations to vectors. $\mathcal{\psi}_{\theta}^R=\mathcal{E}_{\theta} \circ \mathcal{R}$ denotes the complete \textit{encoding}: the combination of the \textit{representation} and \textit{encoder}, thus mapping RNAs to vectors.
The symmetric notations with $L$ superscript apply to ligand representations, encoders and encodings.

\subsection{RNA encodings}
\label{supp:RNA_encodings}

\begingroup
\setlength{\arrayrulewidth}{0.3pt}
\begin{longtable}{@{}llccc@{}}
\toprule
& & \multicolumn{3}{c}{\textbf{\shortstack[c]{RNA encoding\\ $\psi_{\theta}^R$}}} \\
\cmidrule(lr){3-5}
\multicolumn{1}{c}{\textbf{Task}} & \multicolumn{1}{c}{\textbf{Model}} & \multicolumn{1}{c}{\textbf{Structure level}} & \multicolumn{1}{c}{\textbf{\shortstack[c]{Representation \\ $\mathcal{R}^R$}}} & \multicolumn{1}{c}{\textbf{\shortstack[c]{Encoder \\ $\mathcal{E}_\theta^R$}}}\\
\midrule
\endfirsthead

\multicolumn{5}{c}{\textbf{Table \ref{sup:tab:exhaustive}} (Continued)} \\
\toprule
& & \multicolumn{3}{c}{\textbf{\shortstack[c]{RNA encoding\\ $\psi_{\theta}^R$}}} \\
\cmidrule(lr){3-5}
\multicolumn{1}{c}{\textbf{Task}} & \multicolumn{1}{c}{\textbf{Model}} & \multicolumn{1}{c}{\textbf{Structure level}} & \multicolumn{1}{c}{\textbf{\shortstack[c]{Representation \\ $\mathcal{R}^R$}}} & \multicolumn{1}{c}{\textbf{\shortstack[c]{Encoder \\ $\mathcal{E}_\theta^R$}}}\\
\midrule
\endhead

\multirow{19}{*}{\shortstack[c]{Binding site \\ prediction}} & RNASite \cite{rnasite} & Tertiary structure & Expert-based features & \cellcolor[gray]{0.8}\\
& RNACavityMiner \cite{rnacavityminer} & Tertiary structure & Expert-based features & \cellcolor[gray]{0.8}\\
& DrugPredRNA \cite{drugpredrna} & Tertiary structure & Expert-based features & \cellcolor[gray]{0.8}\\
& NABS \cite{nabs} & Tertiary structure & Expert-based features & \cellcolor[gray]{0.8}\\
& RLBind \cite{rlbind} & Tertiary structure & Expert-based features & \cellcolor[gray]{0.8}\\
& RNetsite \cite{rnet} & Tertiary structure & Expert-based features & \cellcolor[gray]{0.8}\\
& ZHmolReSTasite \cite{zhmolrestasite} & Tertiary structure & Expert-based features & \cellcolor[gray]{0.8}\\
& CapBind \cite{capbind} & Tertiary structure & Expert-based features & \cellcolor[gray]{0.8}\\
& DRLiPS \cite{drlips} & Tertiary structure & Expert-based features & \cellcolor[gray]{0.8}\\
& RNABind \cite{rnabind} & Tertiary structure & 3D Nucleotides graph & EGNN \cite{EGNN}\\ 
& \multirow{3}{*}{MultimodRLBP \cite{multimodrlbp}} & Sequence & Sequence & RNABert \cite{rnabert}\\
&& Tertiary structure & Expert-based features & 1D CNN\\
&& Tertiary structure & 2.5D Nucleotides graph & RGCN \cite{rgcn}\\
& RLBSIF \cite{rlbsif} & Tertiary structure & 3D Surface & MoNET \cite{MoNET}\\
& BiteNet \cite{bitenet} & Tertiary structure & 3D Grid & 3D CNN\\
& RNet \cite{moller2022} & Tertiary structure & 3D Grid & 3D CNN\\
& \multirow{2}{*}{RLsite \cite{rlsite}} & Sequence & Sequence & ERNIE-RNA \cite{ernierna}\\
&& Tertiary structure & 2D Nucleotides graph & GAT \cite{gat}\\
& CoBRA \cite{cobra} & Sequence & Sequence & ERNIE-RNA \cite{ernierna}\\
\midrule
\multirow{5}{*}{\centering \shortstack[c]{Binding site\\and affinity\\prediction}} & SMARTBind \cite{smartbind} & Sequence & Sequence & \shortstack[c]{RNA-FM \cite{rnafm}\\(pretrained)}\\
& DeepRNA-DTI \cite{deeprnadti} & Sequence & Sequence & \shortstack[c]{RNA-FM \cite{rnafm}\\(pretrained)}\\
& \multirow{3}{*}{GerNA-Bind \cite{gernabind}} & Sequence & Sequence & \shortstack[c]{RNA-FM \cite{rnafm}\\(pretrained)\\+ MLP}\\
&& Secondary structure & 2D Nucleotides graph & GAT \cite{gat}\\
&& Tertiary structure & 3D Atomic graph & Equiformer \cite{equiformer}\\
\midrule
\pagebreak
\multirow{13}{*}{\shortstack[c]{Binding\\affinity\\prediction}} & RFSMMA \cite{rfsmma} & \cellcolor[gray]{0.8} & Expert-based features & \cellcolor[gray]{0.8}\\
& SMTRS \cite{smtrs} & Sequence & Expert-based features & \cellcolor[gray]{0.8}\\
& SMAJL \cite{smajl} & Sequence & Expert-based features & \cellcolor[gray]{0.8}\\
& RNAincoder \cite{rnaincoder} & Sequence & Expert-based features & \shortstack[c]{Stacked\\autoencoder \cite{pan2016ipminer}}\\
& RSAPred \cite{rsapred} & Sequence & Expert-based features & \cellcolor[gray]{0.8}\\
& ZHMol-RLinter \cite{zhmolrlinter} & Tertiary structure & Expert-based features & \cellcolor[gray]{0.8}\\
& \multirow{2}{*}{EMMPTNet \cite{emmptnet}} & Sequence & Expert-based features & \cellcolor[gray]{0.8}\\
&& Sequence & De Bruijn graph & GCN \cite{gcn}\\
& RNAsmol \cite{rnasmol} & Sequence & Sequence & 1D CNN\\
& BioLLMNet \cite{biollmnet} & Sequence & Sequence & RNA-FM \cite{rnafm}\\
& RLASIF \cite{rlasif} & Tertiary structure & 3D Surface & MoNET \cite{MoNET}\\
& RNAmigos2 \cite{rnamigos2} & Tertiary structure & 2.5D Nucleotides graph & RGCN \cite{rgcn}\\
& \multirow{2}{*}{DeepRSMA \cite{deeprsma}} & Sequence & Sequence & \shortstack[c]{RNA-FM \cite{rnafm}\\(pretrained)\\+ 1D CNN}\\
&& Secondary structure & 2D Nucleotides graph & GAT \cite{gat}\\
\midrule
\multirow{2}{*}{\shortstack[c]{Ligand\\prediction}} & RNAmigos \cite{rnamigos} & Tertiary structure & 2.5D Nucleotides graph & RGCN \cite{rgcn}\\
& E3NN \cite{E3NN} & Tertiary structure & 3D Atomic graph & NequIP \cite{nequip}\\
\bottomrule
\caption{Summary of RNA encodings}
\label{sup:tab:exhaustive}
\end{longtable}
\endgroup

\subsection{Small molecule encodings}
\label{supp:SM_encodings}
\begingroup
\centering
\begin{longtable}[htbp]{@{}clcc@{}}
\toprule
 & & \multicolumn{2}{c}{\textbf{\shortstack[c]{Small molecule encoding\\ $\psi_{\theta}^L$}}} \\
\cmidrule(lr){3-4}
\multicolumn{1}{c}{\textbf{Task}} & \multicolumn{1}{c}{\textbf{Model}} & \multicolumn{1}{c}{\textbf{\shortstack[c]{Representation \\ $\mathcal{R}^L$}}} & \multicolumn{1}{c}{\textbf{\shortstack[c]{Encoder \\ $\mathcal{E}_\theta^L$}}}\\
\midrule
\endfirsthead

\multicolumn{4}{c}{\textbf{Table \ref{sup:tab:small_molecule_encodings}} (Continued)} \\
\toprule
 & & \multicolumn{2}{c}{\textbf{\shortstack[c]{Small molecule encoding\\ $\psi_{\theta}^L$}}} \\
\cmidrule(lr){3-4}
\multicolumn{1}{c}{\textbf{Task}} & \multicolumn{1}{c}{\textbf{Model}} & \multicolumn{1}{c}{\textbf{\shortstack[c]{Representation \\ $\mathcal{R}^L$}}} & \multicolumn{1}{c}{\textbf{\shortstack[c]{Encoder \\ $\mathcal{E}_\theta^L$}}}\\
\midrule
\endhead
\multirow{17}{*}{\shortstack[c]{Binding affinity \\ prediction}} & RFSMMA \cite{rfsmma} & Expert-based features & \cellcolor[gray]{0.8}\\
& SMTRS \cite{smtrs} & MACCS fingerprint & \cellcolor[gray]{0.8}\\
& SMAJL \cite{smajl} & MACCS fingerprint & \cellcolor[gray]{0.8}\\
& RNAincoder \cite{rnaincoder} & Expert-based features & \shortstack[c]{Stacked autoencoder \cite{pan2016ipminer}}\\
& RSAPred \cite{rsapred} & Expert-based features & \cellcolor[gray]{0.8}\\
& SMARTBind \cite{smartbind} & Expert-based features &\cellcolor[gray]{0.8}\\
& ZHMolRLinter \cite{zhmolrlinter} & MACCS fingerprint & \cellcolor[gray]{0.8}\\
& EMMPTNet \cite{emmptnet} & Expert-based features & \cellcolor[gray]{0.8}\\
&& SMILES sequence & BiLSTM\\
& RNAsmol \cite{rnasmol} & 2D Atomic graph & Graph diffusion convolution\\
& RNAmigos2 \cite{rnamigos2} & 2D Atomic graph & OptiMol \cite{optimol}\\
& DeepRSMA \cite{deeprsma} & SMILES sequence & Transformer\\
&& 2D Atomic graph & GCN \cite{gcn}\\
& BioLLMNet \cite{biollmnet} & 3D Atomic graph & Mole-BERT \cite{molebert}\\
& GerNA-Bind \cite{gernabind} & 2D Atomic graph & GCN \cite{gcn}\\
&& 3D Atomic graph & Equiformer \cite{equiformer}\\
& RLASIF \cite{rlasif} & 3D Surface & MoNET \cite{MoNET}\\
& DeepRNA-DTI \cite{deeprnadti} & 3D Atomic graph & Mole-BERT \cite{molebert}\\
\midrule
\multirow{3}{*}{\shortstack[c]{Ligand prediction}} & RNAmigos \cite{rnamigos} & MACCS fingerprint & \cellcolor[gray]{0.8}\\
& E3NN \cite{E3NN} & MACCS fingerprint & \cellcolor[gray]{0.8}\\
&& 3D Atomic graph & Uni-Mol \cite{unimol}\\
\midrule
\multirow{8}{*}{QSAR} & sChemNet \cite{schemnet} & MACCS fingerprint & MLP\\
& Grimberg \textit{et al.}, 2022 \cite{grimberg2022} & MACCS fingerprint & \cellcolor[gray]{0.8}\\
&& SMILES matrix & 2D CNN\\
&& Image & 2D CNN\\
& Cai \textit{et al.}, 2022 \cite{qsar} & Expert-based features & \cellcolor[gray]{0.8}\\
& Rizvi \textit{et al.}, 2020 \cite{rizvi2020targeting} & Expert-based features & \cellcolor[gray]{0.8}\\
& Yazdani \textit{et al.}, 2023 \cite{yazdani2023} & Expert-based features & \cellcolor[gray]{0.8}\\
& Haga \textit{et al.}, 2023 \cite{haga2023} & 3D Atomic graph & GIN \cite{gin}\\
\bottomrule
\caption{Summary of small molecule encodings}
\label{sup:tab:small_molecule_encodings}
\end{longtable}
\endgroup

\subsection{Complex encodings}
\label{supp:complex_encodings}

\begin{table}[htbp]
\centering
\setlength{\arrayrulewidth}{0.3pt}
\begin{tabular}{@{}llcc@{}}
\toprule
 & & \multicolumn{2}{c}{\textbf{\shortstack[c]{Complex encoding\\ $\psi_{\theta}^C$}}} \\
\cmidrule(lr){3-4}
\multicolumn{1}{c}{\textbf{Task}} & \multicolumn{1}{c}{\textbf{Model}} & \multicolumn{1}{c}{\textbf{\shortstack[c]{Representation \\ $\mathcal{R}^C$}}} & \multicolumn{1}{c}{\textbf{\shortstack[c]{Encoder \\ $\mathcal{E}_\theta^C$}}}\\
\midrule
\multirow{8}{*}{Pose scoring} & SIFt-ML \cite{siftml} & Expert-based features & \cellcolor[gray]{0.8}\\
& RNAPosers \cite{rnaposers} & Expert-based features & \cellcolor[gray]{0.8}\\
& AnnapuRNA \cite{annapurna} & Expert-based features & \cellcolor[gray]{0.8}\\
& ssMD \cite{ssmd} & Expert-based features & \cellcolor[gray]{0.8}\\
& RmsdXNA \cite{rmsdxna} & Expert-based features & \cellcolor[gray]{0.8}\\
& IRIS \cite{iris} & Expert-based features & \cellcolor[gray]{0.8}\\
& AffiGrapher \cite{affigrapher} & 3D Atomic graph & Message passing neural network\\
& RLaffinity \cite{rlaffinity} & 3D Grid & 3D CNN\\
\bottomrule
\end{tabular}
\caption{Summary of complex encodings}
\end{table}

\section{Other tasks}
Other applications of machine learning relevant to RNA targeting with small molecules includes the use of machine learning models to enhance molecular dynamics (MD) simulations \cite{mlformd} as well as the use of generative models to generate \textit{holo} conformations of RNAs from \textit{apo} conformations of the same RNAs, as proposed in Molearn \cite{molearn}.

\section{Details about the datasets used}
\label{supp:details_datasets}
The most popular databases for RNA-small molecule dataset construction are the Protein Data Bank (PDB) \cite{pdb} is the main source of data (used in RNASite \cite{rnasite}, RNACavityMiner \cite{rnacavityminer}, DrugPredRNA \cite{drugpredrna}, NABS \cite{nabs}, DRLiPS \cite{drlips}, MultimodRLBP \cite{multimodrlbp}, RLBSIF \cite{rlbsif}, RNAmigos \cite{rnamigos}, RNAmigos 2 \cite{rnamigos2}, DeepRNA-DTI \cite{deeprnadti}, SIFt-ML combination \cite{siftml}, RNAPosers \cite{rnaposers}, AnnapuRNA \cite{annapurna} and RLAffinity \cite{rlaffinity}, and used to construct the dataset introduced in LigandRNA \cite{ligandrna}, itself widely used) and ROBIN \cite{yazdani2023} (used in RNAsmol \cite{rnasmol}, GerNA-Bind \cite{gernabind} and RNAmigos 2 \cite{rnamigos2}). E3NN \cite{E3NN}, RNABind \cite{rnabind} and GerNA-Bind \cite{gernabind} rely on HARIBOSS \cite{hariboss}, a curated dataset of small molecule-binding RNAs. Two databases are specific to feature-based methods: SM2miR, which is specialized in miRNAs (used in RFSMMA \cite{rfsmma}, SMAJL \cite{smajl} and sChemNet \cite{schemnet}) and R-SIM \cite{rsim} (used in RSAPred \cite{rsapred}, DeepRSMA \cite{deeprsma}, BioLLMNet \cite{biollmnet} and Boltz-2 \cite{boltz2}).

\section{Details about the splitting strategies}
RNAPosers \cite{rnaposers}, RNASite \cite{rnasite}, BiteNet \cite{bitenet}, RLBind \cite{rlbind}, CapBind \cite{capbind}, ZHMolReSTasite \cite{zhmolrestasite}, RNet \cite{rnet}, MultimodRLBP \cite{multimodrlbp}, and DeepRNA-DTI \cite{deeprnadti} rely on sequence similarity-based clustering. RNAmigos 2 \cite{rnamigos2}, RNABind \cite{rnabind} and DRLiPS \cite{drlips} implement structural similarity-based clustering.

GerNA-Bind \cite{gernabind}, sChemNet \cite{schemnet}, AnnapuRNA \cite{annapurna}, SPRank \cite{sprank}, and DeepRNA-DTI \cite{deeprnadti} perform clustering based on chemical similarity between ligands and GerNA-Bind \cite{gernabind} relies on temporal splitting for binding site prediction.

\section{Details about figure construction}
Figures \ref{fig:dataset}A-D were constructed by querying the Protein Data Bank (PDB) \cite{pdb}. In Figure \ref{fig:dataset}A, the RNA and protein structures are selected using respectively Number of Distinct RNA Entities $>$ 0 and Number of Distinct Protein Entities $>$ 0. Data reported for each year correspond to the number of structures available on January 1st of the corresponding year. 

In Figure \ref{fig:dataset}B, the RNA-ligand complex structures retained are the structures containing at least 1 RNA entity and 1 Non-polymer entity having a non-polymer entity having a weight between 160 and 1,000 Da (corresponding to the range of drug-like small molecules). 

The pie chart of Figure \ref{fig:dataset}D, RNA structures were classified according to the Nucleic Acids Knowledgebase (NAKB) \cite{nakb} functional classification. Molecules having multiple annotations were excluded from the pie chart. 

For Figures \ref{fig:dataset}E and \ref{fig:dataset}F, we used the data regarding RNA and protein binders from the PDBbind database \cite{pdbbind} and RNA binders from R-SIM database \cite{rsim}. The drug-likeness filter proposed in HARIBOSS \cite{hariboss} was applied to these compounds: we only retained molecules whose weight is between 160 and 1,000 Da and containing at least one C atom and no other atom types than C, H, N, O, Br, Cl, F, P, Si, B, S, Se.
Figure \ref{fig:dataset}E is a t-SNE representation computed from the Morgan chemical fingerprints of small molecules. 

\section{Extensive results of the ablation study}
\label{sec:ablation_study}

In this section, we detail the ablation study. We benchmark six binding affinity prediction models: Boltz-2 \cite{boltz2}, RNAmigos 2 \cite{rnamigos2}, RSAPred \cite{rsapred}, DeepRSMA \cite{deeprsma}, GerNA-Bind \cite{gernabind}, and RNAsmol \cite{rnasmol}. These were selected because their code is open source. They encompass various representations and architectures: RSAPred \cite{rsapred} is a linear regression on expert-based features, Boltz-2 \cite{boltz2} is a sequence-based co-folding model, RNAsmol \cite{rnasmol} is a sequence-based 1D CNN, RNAmigos 2 \cite{rnamigos2} is a graph neural network relying on 2.5D graphs, and DeepRSMA \cite{deeprsma} and GerNA-Bind \cite{gernabind} rely on multimodal representations (sequence and 2D graph for DeepRSMA, and sequence, 2D graph and 3D graph for GerNA-Bind).

\medskip 

We perform inference of each of the models on the datasets they were tested on, except Boltz-2 \cite{boltz2}, which was not tested on RNA-ligand data. We therefore test Boltz-2 on a subset of the ROBIN dataset. More precisely, we select all RNA-ligand couples from ROBIN involving an RNA among SAM\_ll, ZTP, TPP, PreQ1, NRAS, and RRE2B. 

For RNAmigos 2, inference is performed on the ChEMBL dataset. 

For RSAPred, it is performed on the ROBIN dataset (we report performance for riboswitches). 

For DeepRSMA, the inference was performed on the five-fold cross validation dataset extracted from R-SIM \cite{rsim} because the only weights entirely available to us were those of the cross-validation models. 

GerNA-Bind has been tested on both Biosensor \cite{biosensor} and ROBIN \cite{yazdani2023} datasets. The only model with available weights has been trained on the entire dataset, \textit{including the test set}. Therefore, our assessment does not reflect expected model performance in real conditions. Nevertheless, it remains informative for our purpose, which is to decipher whether the information that has been learned by the model is ligand-related only, RNA-related only, or genuinely characterizes RNA-ligand interaction. 

Finally, RNAsmol \cite{rnasmol} was tested both on the PDB dataset extracted by the authors from the Protein Data Bank \cite{pdb}, and the ROBIN dataset. We performed the experiments for the three versions of RNAsmol released by its authors: RNA perturbation, Mol perturbation, and Net perturbation. In RNA perturbation, non-binding samples are generated by adding samples containing the same small molecule as a known sample paired with an RNA whose sequence is a random shuffling of the sequence of the original RNA. In Mol perturbation, non-binding samples are generated by adding samples containing the same RNA as a known sample paired with a small molecule from a distinct chemical library showing high MACCS fingerprint similarity to the original small molecule. In Net perturbation, negative labels are randomly assigned to couples between RNAs and small molecules from the original dataset whose interaction does not have any ground truth label.

\medskip

For each model and dataset tested, we systematically report the performance metrics reported in the corresponding papers. We also compute some additional metrics in order to systematically have results both target-averaged ("target" label in the table) and at the global level ("global" label in the table), that is, averaged across all RNA-small molecule pairs. The interest of target-averaged metrics is to replicate the use of the model in a virtual screening setting, where the model will be used to rank compounds based on their predicted affinity to a specific RNA target. We do not report global results for RNAmigos 2 as it is a ranking model whose output is intrinsically a per-target output.

For each model, we report both results obtained on the dataset where the RNAs have been swapped ("Target swapping" column) and on those where the small molecules have been swapped ("Ligand swapping"). However, the target-averaged metrics are not reported in the Ligand swapping setting (grey boxes in the table). Indeed, since the ligands have been swapped in the test set, the dependency between ligands and labels has been broken. Therefore, it is impossible for the model to correctly rank ligands for a fixed RNA target.

For each swapping experiment, the permutations are randomly generated across three seeds, generating three distinct swapped datasets. Therefore, we report the mean and standard deviation of each performance metric across the three seeds.

\begingroup
\centering
\setlength{\arrayrulewidth}{0.3pt}
\begin{longtable}{@{}lccccc@{}}
\caption{Results of the binding affinity prediction models ablation study (RMSE: Root mean squared error; PCC: Pearson's correlation coefficient; SCC: Spearman's correlation coefficient; AuROC: Area under Receiver-Operating Curve; AuPRC: Area under Precision-Recall Curve)}
\label{ablation_study}
\\
\toprule
 & \multicolumn{2}{c}{\textbf{Metric}} & \multicolumn{1}{c}{\textbf{\shortstack[c]{Without\\swapping}}} & \multicolumn{1}{c}{\textbf{\shortstack[c]{Target\\swapping}}} & \multicolumn{1}{c}{\textbf{\shortstack[c]{Ligand\\swapping}}} \\ 
\midrule
\endfirsthead

\multicolumn{6}{c}{\textbf{Table \ref{ablation_study}} (Continued)} \\
\toprule
 & \multicolumn{2}{c}{\textbf{Metric}} & \multicolumn{1}{c}{\textbf{\shortstack[c]{Without\\swapping}}} & \multicolumn{1}{c}{\textbf{\shortstack[c]{Target\\swapping}}} & \multicolumn{1}{c}{\textbf{\shortstack[c]{Ligand\\swapping}}} \\ 
\midrule
\endhead

\multirow{7}{*}{Boltz-2 \cite{boltz2}} & \multicolumn{1}{l}{AuROC} & \multicolumn{1}{r}{global} & \multicolumn{1}{c}{0.535} & \multicolumn{1}{c}{$0.533 \pm 0.004$} & \multicolumn{1}{c}{$0.506 \pm 0.006$}\\
& & \multicolumn{1}{r}{target} & \multicolumn{1}{c}{0.530} & \multicolumn{1}{c}{$0.529 \pm 0.002$} & \multicolumn{1}{c}{\cellcolor[gray]{0.8}}\\
& \multicolumn{1}{l}{AuPRC} & \multicolumn{1}{r}{global} & \multicolumn{1}{c}{0.009} & \multicolumn{1}{c}{$0.008 \pm 0.000$} & \multicolumn{1}{c}{$0.008 \pm 0.000$}\\
& \multicolumn{1}{l}{Enrichment factor} & \multicolumn{1}{r}{0.5\%} & \multicolumn{1}{c}{1.068} & \multicolumn{1}{c}{$0.649 \pm 0.247$} & \multicolumn{1}{c}{\cellcolor[gray]{0.8}}\\
& & \multicolumn{1}{r}{1\%} & \multicolumn{1}{c}{1.063} & \multicolumn{1}{c}{$1.163 \pm 0.023$} & \multicolumn{1}{c}{\cellcolor[gray]{0.8}}\\
& & \multicolumn{1}{r}{2\%} & \multicolumn{1}{c}{0.988} & \multicolumn{1}{c}{$1.268 \pm 0.078$} & \multicolumn{1}{c}{\cellcolor[gray]{0.8}}\\
& & \multicolumn{1}{r}{5\%} & \multicolumn{1}{c}{1.180} & \multicolumn{1}{c}{$1.102 \pm 0.058$} & \multicolumn{1}{c}{\cellcolor[gray]{0.8}}\\
\midrule
\shortstack[l]{RNAmigos 2 \cite{rnamigos2}\\(PDB decoys)} & \multicolumn{1}{l}{AuROC} & \multicolumn{1}{r}{target} & \multicolumn{1}{c}{0.784} & \multicolumn{1}{c}{$0.580 \pm 0.020$} & \multicolumn{1}{c}{\cellcolor[gray]{0.8}}\\
\midrule
\shortstack[l]{RNAmigos 2 \cite{rnamigos2}\\(Chembl decoys)} & \multicolumn{1}{l}{AuROC} & \multicolumn{1}{r}{target} & \multicolumn{1}{c}{0.899} & \multicolumn{1}{c}{$0.791 \pm 0.007$} & \multicolumn{1}{c}{\cellcolor[gray]{0.8}}\\
\midrule
\multirow{5}{*}{RSAPred \cite{rsapred}} & \multicolumn{1}{l}{AuROC} & \multicolumn{1}{r}{global} & \multicolumn{1}{c}{0.600} & \multicolumn{1}{c}{$0.670 \pm 0.057$} & \multicolumn{1}{c}{$0.566 \pm 0.002$}\\
& & \multicolumn{1}{r}{target} & \multicolumn{1}{c}{0.581} & \multicolumn{1}{c}{$0.449 \pm 0.047$} & \multicolumn{1}{c}{\cellcolor[gray]{0.8}}\\
& \multicolumn{1}{l}{AuPRC} & \multicolumn{1}{r}{target} & \multicolumn{1}{c}{0.352} & \multicolumn{1}{c}{$0.457 \pm 0.017$} & \multicolumn{1}{c}{\cellcolor[gray]{0.8}}\\
& \multicolumn{1}{l}{F1-Score} & \multicolumn{1}{r}{global} & \multicolumn{1}{c}{0.681} & \multicolumn{1}{c}{$0.692 \pm 0.039$} & \multicolumn{1}{c}{$0.651 \pm 0.001$}\\
& & \multicolumn{1}{r}{target} & \multicolumn{1}{c}{0.455} & \multicolumn{1}{c}{$0.474 \pm 0.032$} & \multicolumn{1}{c}{\cellcolor[gray]{0.8}}\\
\midrule
\multirow{6}{*}{DeepRSMA \cite{deeprsma}} & \multicolumn{1}{l}{PCC} & \multicolumn{1}{r}{global} & \multicolumn{1}{c}{0.784} & \multicolumn{1}{c}{$0.424 \pm 0.064$} & \multicolumn{1}{c}{$0.256 \pm 0.022$} \\
& & \multicolumn{1}{r}{target} & \multicolumn{1}{c}{0.430} & \multicolumn{1}{c}{$0.351 \pm 0.021$} & \multicolumn{1}{c}{\cellcolor[gray]{0.8}}\\
& \multicolumn{1}{l}{SCC} & \multicolumn{1}{r}{global} & \multicolumn{1}{c}{0.787} & \multicolumn{1}{c}{$0.407 \pm 0.082$} & \multicolumn{1}{c}{$0.255 \pm 0.026$}\\
& & \multicolumn{1}{r}{target} & \multicolumn{1}{c}{0.405} & \multicolumn{1}{c}{$0.333 \pm 0.022$} & \multicolumn{1}{c}{\cellcolor[gray]{0.8}}\\
& \multicolumn{1}{l}{RMSE} & \multicolumn{1}{r}{global} & \multicolumn{1}{c}{0.895} & \multicolumn{1}{c}{$1.414 \pm 0.060$} & \multicolumn{1}{c}{$1.685 \pm 0.088$}\\
& & \multicolumn{1}{r}{target} & \multicolumn{1}{c}{0.895} & \multicolumn{1}{c}{$1.208 \pm 0.095$} & \multicolumn{1}{c}{\cellcolor[gray]{0.8}}\\
\midrule
\pagebreak
\multirow{5}{*}{\shortstack[l]{GerNA-Bind \cite{gernabind}\\(Biosensor)}} & \multicolumn{1}{l}{AuROC} & \multicolumn{1}{r}{global} & \multicolumn{1}{c}{0.976} & \multicolumn{1}{c}{$0.675 \pm 0.010$} & \multicolumn{1}{c}{$0.632 \pm 0.068$}\\
& & \multicolumn{1}{r}{target} & \multicolumn{1}{c}{0.967} & \multicolumn{1}{c}{$0.728 \pm 0.037$} & \multicolumn{1}{c}{\cellcolor[gray]{0.8}}\\
& \multicolumn{1}{l}{AuPRC} & \multicolumn{1}{r}{target} & \multicolumn{1}{c}{0.599} & \multicolumn{1}{c}{$0.335 \pm 0.028$} & \multicolumn{1}{c}{\cellcolor[gray]{0.8}}\\
& \multicolumn{1}{l}{F1-Score} & \multicolumn{1}{r}{global} & \multicolumn{1}{c}{0.715} & \multicolumn{1}{c}{$0.266 \pm 0.010$} & \multicolumn{1}{c}{$0.217 \pm 0.024$}\\
& & \multicolumn{1}{r}{target} & \multicolumn{1}{c}{0.509} & \multicolumn{1}{c}{$0.178 \pm 0.003$} & \multicolumn{1}{c}{\cellcolor[gray]{0.8}}\\
\midrule
\multirow{5}{*}{\shortstack[l]{GerNA-Bind \cite{gernabind}\\(ROBIN)}} & \multicolumn{1}{l}{AuROC} & \multicolumn{1}{r}{global} & \multicolumn{1}{c}{0.783} & \multicolumn{1}{c}{$0.514 \pm 0.005$} & \multicolumn{1}{c}{$0.557 \pm 0.002$}\\
& & \multicolumn{1}{r}{target} & \multicolumn{1}{c}{0.765} & \multicolumn{1}{c}{$0.529 \pm 0.005$} & \multicolumn{1}{c}{\cellcolor[gray]{0.8}}\\
& \multicolumn{1}{l}{AuPRC} & \multicolumn{1}{r}{target} & \multicolumn{1}{c}{0.232} & \multicolumn{1}{c}{$0.105 \pm 0.003$} & \multicolumn{1}{c}{\cellcolor[gray]{0.8}}\\
& \multicolumn{1}{l}{F1-Score} & \multicolumn{1}{r}{global} & \multicolumn{1}{c}{0.264} & \multicolumn{1}{c}{$0.128 \pm 0.002$} & \multicolumn{1}{c}{$0.161 \pm 0.003$}\\
& & \multicolumn{1}{r}{target} & \multicolumn{1}{c}{0.246} & \multicolumn{1}{c}{$0.117 \pm 0.002$} & \multicolumn{1}{c}{\cellcolor[gray]{0.8}}\\
\midrule
\multirow{5}{*}{\shortstack[l]{RNAsmol \cite{rnasmol}\\(PDB)\\Mol perturbation}} & \multicolumn{1}{l}{AuROC} & \multicolumn{1}{r}{global} & \multicolumn{1}{c}{0.983} & \multicolumn{1}{c}{$0.986 \pm 0.005$} & \multicolumn{1}{c}{$0.504 \pm 0.052$}\\
& & \multicolumn{1}{r}{target} & \multicolumn{1}{c}{0.950} & \multicolumn{1}{c}{$0.997 \pm 0.003$} & \multicolumn{1}{c}{\cellcolor[gray]{0.8}}\\
& \multicolumn{1}{l}{AuPRC} & \multicolumn{1}{r}{target} & \multicolumn{1}{c}{0.886} & \multicolumn{1}{c}{$0.789 \pm 0.032$} & \multicolumn{1}{c}{\cellcolor[gray]{0.8}}\\
& \multicolumn{1}{l}{F1-Score} & \multicolumn{1}{r}{global} & \multicolumn{1}{c}{0.935} & \multicolumn{1}{c}{$0.930 \pm 0.014$} & \multicolumn{1}{c}{$0.543 \pm 0.031$}\\
& & \multicolumn{1}{r}{target} & \multicolumn{1}{c}{0.857} & \multicolumn{1}{c}{$0.761 \pm 0.025$} & \multicolumn{1}{c}{\cellcolor[gray]{0.8}}\\
\midrule
\multirow{5}{*}{\shortstack[l]{RNAsmol \cite{rnasmol}\\(PDB)\\RNA perturbation}} & \multicolumn{1}{l}{AuROC} & \multicolumn{1}{r}{global} & \multicolumn{1}{c}{0.991} & \multicolumn{1}{c}{$0.480 \pm 0.038$} & \multicolumn{1}{c}{$0.990 \pm 0.001$}\\
& & \multicolumn{1}{r}{target} & \multicolumn{1}{c}{0.000} & \multicolumn{1}{c}{$0.354 \pm 0.180$} & \multicolumn{1}{c}{\cellcolor[gray]{0.8}}\\
& \multicolumn{1}{l}{AuPRC} & \multicolumn{1}{r}{target} & \multicolumn{1}{c}{0.955} & \multicolumn{1}{c}{$0.449 \pm 0.179$} & \multicolumn{1}{c}{\cellcolor[gray]{0.8}}\\
& \multicolumn{1}{l}{F1-Score} & \multicolumn{1}{r}{global} & \multicolumn{1}{c}{0.962} & \multicolumn{1}{c}{$0.441 \pm 0.016$} & \multicolumn{1}{c}{$0.960 \pm 0.010$}\\
& & \multicolumn{1}{r}{target} & \multicolumn{1}{c}{0.924} & \multicolumn{1}{c}{$0.429 \pm 0.127$} & \multicolumn{1}{c}{\cellcolor[gray]{0.8}}\\
\midrule
\multirow{5}{*}{\shortstack[l]{RNAsmol \cite{rnasmol}\\(PDB)\\Net perturbation}} & \multicolumn{1}{l}{AuROC} & \multicolumn{1}{r}{global} & \multicolumn{1}{c}{0.700} & \multicolumn{1}{c}{$0.575 \pm 0.052$} & \multicolumn{1}{c}{$0.539 \pm 0.078$}\\
& & \multicolumn{1}{r}{target} & \multicolumn{1}{c}{0.723} & \multicolumn{1}{c}{$0.615 \pm 0.059$} & \multicolumn{1}{c}{\cellcolor[gray]{0.8}}\\
& \multicolumn{1}{l}{AuPRC} & \multicolumn{1}{r}{target} & \multicolumn{1}{c}{0.593} & \multicolumn{1}{c}{$0.563 \pm 0.028$} & \multicolumn{1}{c}{\cellcolor[gray]{0.8}}\\
& \multicolumn{1}{l}{F1-Score} & \multicolumn{1}{r}{global} & \multicolumn{1}{c}{0.649} & \multicolumn{1}{c}{$0.518 \pm 0.043$} & \multicolumn{1}{c}{$0.495 \pm 0.083$}\\
& & \multicolumn{1}{r}{target} & \multicolumn{1}{c}{0.497} & \multicolumn{1}{c}{$0.335 \pm 0.023$} & \multicolumn{1}{c}{\cellcolor[gray]{0.8}}\\
\midrule
\multirow{5}{*}{\shortstack[l]{RNAsmol \cite{rnasmol}\\(ROBIN)\\Mol perturbation}} & \multicolumn{1}{l}{AuROC} & \multicolumn{1}{r}{global} & \multicolumn{1}{c}{0.974} & \multicolumn{1}{c}{$0.975 \pm 0.000$} & \multicolumn{1}{c}{$0.486 \pm 0.018$}\\
& & \multicolumn{1}{r}{target} & \multicolumn{1}{c}{0.972} & \multicolumn{1}{c}{$0.973 \pm 0.002$} & \multicolumn{1}{c}{\cellcolor[gray]{0.8}}\\
& \multicolumn{1}{l}{AuPRC} & \multicolumn{1}{r}{target} & \multicolumn{1}{c}{0.968} & \multicolumn{1}{c}{$0.974 \pm 0.002$} & \multicolumn{1}{c}{\cellcolor[gray]{0.8}}\\
& \multicolumn{1}{l}{F1-Score} & \multicolumn{1}{r}{global} & \multicolumn{1}{c}{0.898} & \multicolumn{1}{c}{$0.898 \pm 0.002$} & \multicolumn{1}{c}{$0.546 \pm 0.012$}\\
& & \multicolumn{1}{r}{target} & \multicolumn{1}{c}{0.889} & \multicolumn{1}{c}{$0.897 \pm 0.003$} & \multicolumn{1}{c}{\cellcolor[gray]{0.8}}\\
\midrule
\multirow{5}{*}{\shortstack[l]{RNAsmol \cite{rnasmol}\\(ROBIN)\\RNA perturbation}} & \multicolumn{1}{l}{AuROC} & \multicolumn{1}{r}{global} & \multicolumn{1}{c}{1.000} & \multicolumn{1}{c}{$0.511 \pm 0.024$} & \multicolumn{1}{c}{$1.000 \pm 0.000$}\\
& & \multicolumn{1}{r}{target} & \multicolumn{1}{c}{0.500} & \multicolumn{1}{c}{$0.506 \pm 0.062$} & \multicolumn{1}{c}{\cellcolor[gray]{0.8}}\\
& \multicolumn{1}{l}{AuPRC} & \multicolumn{1}{r}{target} & \multicolumn{1}{c}{0.710} & \multicolumn{1}{c}{$0.576 \pm 0.040$} & \multicolumn{1}{c}{\cellcolor[gray]{0.8}}\\
& \multicolumn{1}{l}{F1-Score} & \multicolumn{1}{r}{global} & \multicolumn{1}{c}{0.999} & \multicolumn{1}{c}{$0.521 \pm 0.019$} & \multicolumn{1}{c}{$0.999 \pm 0.000$}\\
& & \multicolumn{1}{r}{target} & \multicolumn{1}{c}{0.710} & \multicolumn{1}{c}{$0.479 \pm 0.014$} & \multicolumn{1}{c}{\cellcolor[gray]{0.8}}\\
\midrule
\multirow{5}{*}{\shortstack[l]{RNAsmol \cite{rnasmol}\\(ROBIN)\\Net perturbation}} & \multicolumn{1}{l}{AuROC} & \multicolumn{1}{r}{global} & \multicolumn{1}{c}{0.717} & \multicolumn{1}{c}{$0.546 \pm 0.009$} & \multicolumn{1}{c}{$0.604 \pm 0.012$}\\
& & \multicolumn{1}{r}{target} & \multicolumn{1}{c}{0.651} & \multicolumn{1}{c}{$0.569 \pm 0.008$} & \multicolumn{1}{c}{\cellcolor[gray]{0.8}}\\
& \multicolumn{1}{l}{AuPRC} & \multicolumn{1}{r}{target} & \multicolumn{1}{c}{0.669} & \multicolumn{1}{c}{$0.615 \pm 0.003$} & \multicolumn{1}{c}{\cellcolor[gray]{0.8}}\\
& \multicolumn{1}{l}{F1-Score} & \multicolumn{1}{r}{global} & \multicolumn{1}{c}{0.062} & \multicolumn{1}{c}{$0.038 \pm 0.011$} & \multicolumn{1}{c}{$0.038 \pm 0.006$}\\
& & \multicolumn{1}{r}{target} & \multicolumn{1}{c}{0.054} & \multicolumn{1}{c}{$0.033 \pm 0.009$} & \multicolumn{1}{c}{\cellcolor[gray]{0.8}}\\
\bottomrule
\end{longtable}
\endgroup

\end{document}